\documentclass{aa}

\pdfoutput=1

\newcommand{\bol}{\boldsymbol}
\newcommand{\btheta}{\ensuremath{\boldsymbol{\theta}}}
\newcommand{\eq}[1]{\begin{align}#1\end{align}}

\newcommand{\dd}{\mathrm{d}}

\newcommand{\gt}{\ensuremath{\gamma_\mathrm{t}}}
\newcommand{\tgt}{\ensuremath{\delta_\gamma}}
\newcommand{\dM}{\ensuremath{\delta_M}}
\newcommand{\gx}{\ensuremath{\gamma_\times}}
\newcommand{\gtx}{\ensuremath{\gamma_{\mathrm{t}/\times}}}

\newcommand{\Ds}{\ensuremath{D_\mathrm{s}}}
\newcommand{\Dds}{\ensuremath{D_\mathrm{ds}}}
\newcommand{\Dd}{\ensuremath{D_\mathrm{d}}}
\newcommand{\zd}{\ensuremath{z_\mathrm{d}}}
\newcommand{\zs}{\ensuremath{z_\mathrm{s}}}
\newcommand{\zds}{\ensuremath{z_\mathrm{d,s}}}
\newcommand{\ad}{\ensuremath{\alpha_\mathrm{d}}}
\newcommand{\as}{\ensuremath{\alpha_\mathrm{s}}}
\newcommand{\ads}{\ensuremath{\alpha_\mathrm{d,s}}}
\newcommand{\ns}{\ensuremath{n_\mathrm{s}}}
\newcommand{\nd}{\ensuremath{n_\mathrm{d}}}
\newcommand{\slims}{\ensuremath{s_\mathrm{lim,s}}}
\newcommand{\slimd}{\ensuremath{s_\mathrm{lim,d}}}
\newcommand{\Msun}{\ensuremath{h^{-1} \mathrm{M}_\odot}}

\newcommand{\gs}{\gamma_{\mathrm{s}}}

\newcommand{\pzs}{p_{\zs}}
\newcommand{\pzd}{p_{\zd}}
\newcommand{\pmforzd}{p_{m|\zd}}

\newcommand{\deltaK}{\delta_{\mathrm{K}}}

\newcommand{\degt}{\mathrm{deg}}

\newcommand{\arcsect}{\mathrm{arcsec}}

\newcommand{\est}[1]{\hat{#1}}
\newcommand{\ev}[1]{\left\langle #1 \right\rangle}
\newcommand{\ft}[1]{\mathcal{F}\left\{#1\right\}}
\newcommand{\ift}[1]{\mathcal{F}^{-1}\left\{#1\right\}}

\usepackage{amsmath, amssymb}
\usepackage[english=british]{csquotes}
\usepackage[T1]{fontenc}
\usepackage[utf8]{inputenc}
\usepackage{lmodern}
\usepackage{graphicx}
\usepackage[dvipsnames]{xcolor}
\usepackage{txfonts}
\usepackage{natbib}
 \bibpunct{(}{)}{;}{a}{}{,}
\usepackage{hyperref}

\usepackage{ulem}

\makeatletter
\renewcommand*\aa@pageof{, page \thepage{} of \pageref*{LastPage}}
\makeatother

\begin{document} 
\title{The importance of magnification effects in galaxy-galaxy lensing}
\author{Sandra Unruh\inst{1}, Peter Schneider\inst{1}, Stefan Hilbert\inst{2,3}, Patrick Simon\inst{1}, Sandra Martin\inst{1} \and Jorge Corella Puertas\inst{2,4}}
\institute{
    Argelander-Institut f\"ur Astronomie, Universit\"at Bonn, Auf dem H\"ugel 71, D-53121 Bonn, Germany\\     sandra@astro.uni-bonn.de
    \and
    Exzellenzcluster Universe, Boltzmannstr. 2, D-85748 Garching, Germany
    \and
    Ludwig-Maximilians-Universit{\"a}t, Universit{\"a}ts-Sternwarte, Scheinerstr. 1, D-81679 M{\"u}nchen, Germany
    \and
    Physik-Department, Technische Universit{\"a}t M{\"u}nchen, 85748 Garching, Germany
}
\date{Version \today; received Oct.19, accepted yyy} 
\abstract{Magnification changes the observed local number density of galaxies on the sky. This biases the observed tangential shear profiles around galaxies: the so-called galaxy-galaxy lensing (GGL) signal. Inference of physical quantities, such as the mean mass profile of halos around galaxies, are correspondingly affected by magnification effects. We used simulated shear and galaxy data from the Millennium Simulation to quantify the effect on shear and mass estimates from the magnified lens and source number counts. The former is due to the large-scale matter distribution in the foreground of the lenses; the latter is caused by magnification of the source population by the matter associated with the lenses. The GGL signal is calculated from the simulations by an efficient fast Fourier transform, which can also be applied to real data. The numerical treatment is complemented by a leading-order analytical description of the magnification effects, which is shown to fit the numerical shear data well. We find the magnification effect is strongest for steep galaxy luminosity functions and high redshifts. For a KiDS+VIKING+GAMA-like survey with lens galaxies at redshift \(\zd=0.36\) and source galaxies in the last three redshift bins with a mean redshift of \(\bar{z}_\mathrm{s}=0.79\), the magnification correction changes the shear profile up to \(2\%,\) and the mass is biased by up to \(8 \%\). We further considered an even higher redshift fiducial lens sample at \(\zd=0.83,\) with a limited magnitude of \(22\,\mathrm{mag}\) in the \(r\)-band and a source redshift of \(\zs=0.99\). Through this, we find that a magnification correction changes the shear profile up to \(45\%\) and that the mass is biased by up to \(55 \%\). As expected, the sign of the bias depends on the local slope of the lens luminosity function \ad, where the mass is biased low for \(\ad<1\) and biased high for \(\ad>1\). While the magnification effect of sources is rarely more than \(1\%\) of the measured GGL signal, the statistical power of future weak lensing surveys warrants correction for this effect.}
%
\titlerunning{Magnification effects in GGL} 
\authorrunning{Unruh et al.}
\maketitle
%
%
\section{Introduction}
\label{sec:introduction}

Gravitational lensing is a powerful tool in unveiling the true distribution of matter in the Universe and probing cosmological parameters \citep[see, e.g.][for a recent review]{Kilbinger2015}. The lensing signal is sensitive to all matter, regardless of its nature, and is observed as the distortion of light bundles travelling through the Universe. In the weak lensing regime, this distortion is small and must be studied with large statistical samples. Therefore, large and deep surveys are required, for example, the Kilo Degree Survey\footnote{\href{http://kids.strw.leidenuniv.nl}{\texttt{https://www.kids.strw.leidenuniv.nl}}}(KiDS); the Dark Energy Survey\footnote{\href{https://www.darkenergysurvey.org}{\texttt{https://www.darkenergysurvey.org}}} (DES); the Hyper Suprime-Cam Subaru Strategic Program\footnote{\href{https://hsc.mtk.nao.ac.jp/ssp/}{\texttt{https://hsc.mtk.nao.ac.jp/ssp/}}} (HSC SSP); or the near-future surveys with \textit{Euclid}\footnote{\href{https://www.euclid-ec.org}{\texttt{https://www.euclid-ec.org}}} and the Large Synoptic Survey Telescope\footnote{\href{https://www.lsst.org}{\texttt{https://www.lsst.org}}} (LSST). To maximise the scientific output from these surveys, the scientific community is currently putting great efforts into understanding nuances in the theoretical framework.

Galaxy-galaxy lensing (GGL) correlates the position of foreground galaxies to the distortion of background galaxies \citep[see, e.g.][]{Hoekstra2013}. The distortion is typically measured in terms of mean tangential shear with respect to the lens position. This shear signal, as a function of separation from the lens centre, can be related to the underlying mass properties of the parent halo. A major challenge is obtaining an unbiased mass estimate. Biases arise if the underlying model does not describe all contributions to the matter-shear correlation function sufficiently. Also, the galaxy bias that connects the position of galaxies to its surrounding matter distribution must be carefully taken into account. For current and future surveys, we must further consider second-order effects to the galaxy-matter correlation, such as, for example, magnification effects \citep{2008PhRvD..78l3517Z,Hilbert2009} and intrinsic alignment of galaxies \citep{Troxel2015}. In this work, we focus on the former effect. 

Magnification is the change of the observed solid angle of an image compared to the intrinsic solid angle, or, since the surface brightness remains constant, the ratio of observed flux to the intrinsic one. Like the shear, it is a local quantity, a direct prediction of the lensing formalism, and is caused by all matter between the observed galaxy population and us. However, direct measurements of magnification are challenging because the intrinsic flux is typically unknown. Yet, the change in size and magnitude results in a changed spatial distribution of the galaxy population. This so-called number count magnification has been measured \citep[e.g.][]{Chiu2016,Garcia-Fernandez2018}. Consequently, magnification by the large-scale structure (LSS) also changes the GGL signal compared to a signal that is just given by matter correlated with the lens galaxies. We stress that the magnification changes the number counts of the source as well as lens galaxies on the sky. The impact of magnification of the lens galaxies on the GGL signal for surveys like CFHTLenS is \(\sim 5\%\) \citep{Simon2018}, but can be as large as \(20\%\) for other lens samples \citep{Hilbert2009}. Although these results suggest a fairly large impact of magnification on GGL lensing estimates, quantitative analyses widely neglect the influence of magnification. \citet{Unruh2019} studied the impact of the number count magnification of lens galaxies on the shear-ratio. They found that the shear-ratio test \citep{jain2003} is affected by lens magnification and that its effect must be mitigated, especially for high-lens redshifts.

In this paper, we quantify the impact of magnification on observed tangential shear profiles and halo-mass estimates from GGL. We consider both the effect of magnification of the sources by the lenses, as well as the effect of magnification of the lenses by the LSS. For this, we compared the GGL signal with and without magnification using simulated data. We then derived mean halo masses in both cases employing a halo model to quantify the expected mass bias. We complemented the numerical results with analytic estimates of the effects.

This article is organised as follows. The theoretical framework is briefly described in Sect.\,\ref{sec:theory}. Section \ref{sec:magneffects} features an analytical description of how magnification affects shear estimates, as well as a brief discussion of how numerical results were obtained in this study. The numerical procedure is then more thoroughly explained in Appendix \ref{sec:mockdata}. In Sect.~\ref{sec:magnification_effects_on_background_galaxies}, we discuss the impact of magnification of source galaxies by the lenses, and in Sect.\,\ref{sec:magnification_effects_on_foreground_galaxies} we discuss magnification of the lens galaxies by the LSS. We study the magnification bias on mass estimates in Sect.\,\ref{sec:magnification_bias_in_halo_mass_estimates} and conclude in Sect.\,\ref{sec:Sc7}.
\section{Theory}
\label{sec:theory}

In the following, we introduce the theoretical concepts of gravitational lensing for this work. For a more general and extensive overview, the reader is kindly referred to \citet{Bartelmann2001}.

\subsection{Cosmological distances}

For a flat universe, the Hubble parameter \(H(z\)) can be written as
\eq{
    \left( \frac{H(z)}{H_0} \right)^2 = \Omega_\mathrm{m} (1+z)^3 + \, 1 - \Omega_\mathrm{m} \;,
}
where \(z\) is the redshift, \(H_0\) denotes the Hubble constant, and \(\Omega_\mathrm{m}\) is the matter density in units of today's critical density \(\rho_\mathrm{crit} = 3H_0^2/(8\pi G)\); with the vacuum speed of light \(c\) and the gravitational constant \(G\). The comoving distance travelled by a photon between redshift \(z_1\) and \(z_2\) reads
\eq{
    \chi(z_1,z_2) = \int^{z_2}_{z_1} \frac{c \, \mathrm{d} z'}{H(z')} \;,
}
and the angular-diameter distance is
\eq{
    D(z_1,z_2) = \frac{\chi(z_1,z_2)}{1+z_2} \;.
}
For a redshift \(z_1=0\), which is the observer's position, we write \(D(0, z) =: D(z)\). In addition, the dimensionless Hubble parameter \(h\) is used to parametrise our ignorance about the true value of today's Hubble parameter, defined as \(H_0 = 100\, h \, \mathrm{km \,s^{-1} \, Mpc^{-1}}\). In the following, all distances are angular-diameter distances.

\subsection{Gravitational lensing distortions and magnification}
\label{ssec:magneffects}

Gravitational lensing distorts the appearance of galaxy images. In the weak lensing regime, this distortion can locally be described as a linear mapping from the background (source) plane to the foreground (lens) plane. The Jacobian \(\mathcal{A}\) of the local mapping can be written as
\eq{
    \mathcal{A} &= \begin{pmatrix} 1-\kappa-\gamma_1 & -\gamma_2 \\ -\gamma_2 & 1-\kappa+\gamma_1 \end{pmatrix}
    = (1-\kappa)\,\begin{pmatrix} 1-g_1 & -g_2 \\ -g_2 & 1+g_1 \end{pmatrix} \;,
}
where \(\kappa\) is the convergence, \(\gamma_{1,2}\) are the two Cartesian shear components, and \(g_{1,2} = \gamma_{1,2}/(1-\kappa)\) are the two Cartesian reduced shear components, which all depend on the position in the lens plane. The convergence causes an isotropic scaling of the galaxy image, while the shear leads to an anisotropic stretching, and thus causes an initially circular object to appear elliptical.

This scaling of the galaxy image changes the apparent solid angle \(\omega\) of the image, compared to one in the absence of lensing, which we denote by \(\omega_0\). Likewise, the flux is affected by gravitational lensing, the unlensed flux \(s_0\) is enhanced or reduced to the observed flux \(s\). The ratio of these quantities defines the magnification \(\mu\) and can also be calculated from the Jacobian by
\eq{
    \mu = \frac{\omega}{\omega_0} = \frac{s}{s_0} = \frac{1}{\det \mathcal{A}} = \frac{1}{(1-\kappa)^2 - |\gamma|^2} \;.
}
%
Magnification changes the observed local number density of galaxies on the sky. The cumulative observed number density of galaxies on the sky \(n(>s)\), brighter than flux \(s\), is locally
\eq{
    \label{eq:cumnumdens}
    n(>s) = \frac{1}{\mu} \, n_0 \left( >\frac{s}{\mu} \right) \;,
}
where \(n_0\) denotes the cumulative number density in absence of lensing. The prefactor \(1/\mu\) is due to the scaling of the solid angle. The flux in the argument of \(n_0\) must also be scaled by \(1/\mu\) to account for the flux enhancement or reduction.

Magnification effects in the weak lensing limit are small, specifically \(|\mu-1|\ll1\), and we Taylor expand Eq.\,(\ref{eq:cumnumdens}) in $(\mu-1)$ to obtain to first order
\eq{
    \label{eq:numchange}
    \frac{n(>s)}{n_0(>s)} = \mu^{\alpha-1.} \;
}
Where the exponent \(\alpha\) is the local slope at the flux limit \(s_\mathrm{lim}\), it is defined as
\eq{
    \label{eq:faintendslope}
    \alpha = - \frac{\mathrm{d} \log_{10} n_0(>s)}{\mathrm{d} \log_{10} s} \Biggm|_{s_\mathrm{lim}} \,.
}
For \(\alpha>1,\) the galaxy counts are enhanced, and for \(\alpha<1\) they are depleted. In the case of \(\alpha=1,\) no magnification bias is present.

\subsection{Galaxy-galaxy lensing}

In GGL, the positions of foreground galaxies (lenses) are correlated with the shear of background galaxies (sources). For a position~\btheta, the complex shear is written as \(\gamma(\btheta) = \gamma_1(\btheta) + \mathrm{i} \gamma_2(\btheta) \).  The tangential shear \gt\ and the cross shear \gx\ at source position \(\btheta_{\mathrm{s}}\) for a given lens at position \(\btheta_\mathrm{d}\) are
\eq{
    \label{eq:tangshear}
    \gt (\btheta_{\mathrm{s}}; \btheta_{\mathrm{s}} - \btheta_\mathrm{d})
                \,+\, \mathrm{i} \gx (\btheta_{\mathrm{s}}; \btheta_{\mathrm{s}} - \btheta_\mathrm{d}) = 
                -\gamma (\btheta_{\mathrm{s}}) \,
                \frac{(\btheta_{\mathrm{s}} - \btheta_\mathrm{d})^*}{\btheta_{\mathrm{s}} - \btheta_\mathrm{d}} \;,
}
where an asterisk denotes complex conjugation. The GGL signal $\langle \gt\rangle(\theta)$ is defined as the correlator between the positions of foreground galaxies and the tangential shear,
\begin{equation}
\langle \gt\rangle(\theta)=\langle \kappa_\mathrm{g}(\btheta')
\;\gt(\btheta'+\btheta;\btheta) \rangle\;,
\label{eq:101}
\end{equation}
where $\kappa_\mathrm{g}(\btheta)$ is the fractional number-density contrast of foreground lens galaxies on the sky. The corresponding correlator for the cross-component of the shear is expected to vanish, due to parity invariance.

A practical estimator for the GGL signal averages the tangential and cross shear over many lens-source pairs in bins of separation $\theta$:
\eq{
    \label{eq:ggl_gtx_estimator}
    \est{\gamma}_{\mathrm{t}/\times} (\theta) =
    \frac
    {
    \sum_{ij}\, \Delta\bigl(\theta, |\btheta_{\mathrm{s}}^{(i)} - \btheta_{\mathrm{d}}^{(j)}|\bigr) \;
    \gtx \bigl(\btheta_{\mathrm{s}}^{(i)}; \btheta_{\mathrm{s}}^{(i)} - \btheta_\mathrm{d}^{(j)}\bigr)
    }
    {\sum_{ij}\, \Delta\bigl(\theta, |\btheta_{\mathrm{s}}^{(i)} - \btheta_{\mathrm{d}}^{(j)}|\bigr)} \;.
}
Here, $\btheta_{\mathrm{s}}^{(i)}$ denotes the position of the $i$-th source, $\btheta_{\mathrm{d}}^{(j)}$ denotes the position of the $j$-th lens, and the binning function $\Delta(\theta, \theta')$ is unity if $\theta'$ falls into the corresponding $\theta$ bin, and zero if it does not.

\section{Magnification effects in GGL}
\label{sec:magneffects}
In this section, we consider the effect of magnification on the GGL signal. As we show, magnification of sources and lenses leads to a bias of the estimator (\ref{eq:ggl_gtx_estimator}), which is a function of limiting magnitudes for the lens and source population, as well as their redshifts, since these determine the local slope \eqref{eq:faintendslope} at the limiting magnitude.  Magnification is typically assumed to be a minor effect in GGL measurements, and most theoretical predictions do not account for it. While the impact of the magnification of lenses has already received some attention \citep[e.g.][]{2008PhRvD..78l3517Z,2009PhDThesisHartlap}, the source magnification is less well known.

\subsection{Magnification of lenses by large-scale structure}
\label{Sect:MlLSS}
Magnification, caused by the LSS between us and the lenses, changes the number density of the lens galaxy sample, while simultaneously inducing a shear on background galaxies. Thus, the observed shear signal differs from what is typically considered as the GGL signal, which is a correlation of lens galaxy positions and the shear on background galaxies (Eq.\,\ref{eq:ggl_gtx_estimator}). In this correlation, the lens galaxies are connected by the galaxy bias to their surrounding matter, which induces a shear in the background galaxies. Magnification by intervening matter structures alters this rather simple picture. Since for larger lens redshifts more intervening matter is present, the impact of magnification effects grows with increasing redshift. On the other hand, the impact is reduced with increasing line-of-sight separations of lenses and sources. In the following, we consider a lowest-order correction for the magnification of the lenses by the LSS for the GGL signal of a flux- or volume-limited lens sample \citep[see, e.g.][]{2008PhRvD..78l3517Z,2009PhDThesisHartlap,Thiele2020}. We stress that this correction ignores the magnification of sources, which is treated in the next sub-section.

In the weak lensing regime, we can approximate the magnification by \(\mu \approx 1+2\kappa\), valid if $\kappa\ll 1$, $|\gamma|\ll 1$. Then, the number count magnification (\ref{eq:numchange}) of the observed number density of lenses $\nd(\bol{\theta})$ at redshift \zd\ on the sky is, for a flux-limited sample,
\begin{equation}
    \label{eq:lensnumcountmagn}
    \nd(\bol{\theta}, \zd) = n_\mathrm{d,0}(\bol{\theta}, \zd) + 2\left[\ad(\zd) - 1 \right]\,\kappa^\mathrm{LSS}(\bol{\theta}, \zd)\,\bar{n}_\mathrm{d}(\zd) \;,
\end{equation}
where $n_\mathrm{d,0}$ denotes the lens number density without magnification, $\bar{n}_\mathrm{d}$ denotes the mean lens number density, $\ad$ denotes the local slope of the lenses at the limiting magnitude, and \(\kappa^\mathrm{LSS}\) denotes the convergence due to matter structures between us and the lenses. Thus, in the presence of magnification, the expected signal is modified to
\begin{align}
    \begin{split}
        \label{eq:lensmagn}
        \gt(\theta| \zd, \zs) 
                &=\gt^\mathrm{nomagn}(\theta| \zd, \zs)
                \\&\qquad
                + 2\left[\ad(\zd) - 1 \right]\,\gt^\mathrm{LSS}(\theta| \zd, \zs) \;,
    \end{split}
\end{align}
where $\gt^\mathrm{nomagn}$ denotes the tangential shear signal without magnification, and the LSS shear signal is
\begin{multline}
\label{eq:ggl_}
    \gt^\mathrm{LSS}(\theta| \zd, \zs) = \frac{9H_0^3\Omega_\mathrm{m}^2}{8 \pi c^3}\,
    \int_0^\infty\mathrm{d}\ell\;\ell\,\mathrm{J}_2(\ell\theta)\,
    \int_0^{\zd}\mathrm{d} z \; (1+z)^2 \frac{H_0}{H(z)}
                \\\times\
    \frac{D(z,\zd)\,D(z,\zs)}{\Dd\,\Ds}\, P_\mathrm{m}\left(\frac{\ell+1/2}{(1+z)\,D(z)};z\right)\;,
\end{multline}
which is shown explicitly in \cite{2009PhDThesisHartlap} and \cite{Simon2018}. We set $D(\zs) = \Ds$ and $D(\zd) = \Dd,$ and by $\mathrm{J}_n(x)$ we denote the $n$th-order Bessel function of the first kind. In this work, we use the revised \texttt{Halofit} model \citep{2012ApJ...761..152T} for the spatial matter power-spectrum $P_\mathrm{m}(k;z)$ at wavenumber $k$ and redshift $z$. The argument in the matter power spectrum arises through the application of the wide-angle corrected Limber projection, which was recently put forward by \cite{2017MNRAS.472.2126K}; the denominator is the comoving angular-diameter distance $f_k(z)=(1+z)D(z)$ at redshift $z$. The corresponding expression of Eq.\,\eqref{eq:lensmagn} in the absence of a flux limit, meaning for a volume-limited lens sample, can be obtained by setting \mbox{$\ad=0$}.

For $\ad < 1$, the magnification by the LSS suppresses the GGL signal. For $\ad = 1$, the magnification effect vanishes. For $\ad > 1$, the LSS contribution enhances the GGL signal.

\subsection{Magnification of sources by lenses}
\label{Sect:Msl}
Galaxies are correlated with the mass distribution, and thus the location of the galaxies correlates with the magnification induced on the background sources. This implies that the number density of sources is correlated with the positions of the lens galaxies. Assuming for a moment that the number-count slope of sources $\alpha$ is larger than unity, one expects that the number density of sources is more enhanced close to lens galaxies living in a dense environment. The estimator (\ref{eq:ggl_gtx_estimator}) therefore contains a disproportionately high number of lens-source pairs for those lenses living in a dense environment compared to those located in less dense regions.

The expected number density of sources is
\begin{align}
  \ns(\btheta, >s)&={\frac{1}{\mu(\btheta)}} \,n_\mathrm{s0}
  \left(>{\frac{s}{\mu(\btheta)}}\right)
  \approx
  n_\mathrm{s0}(>s)\,\mu^{\as-1}(\btheta) \nonumber \\
  &\approx
  n_\mathrm{s0}(>s)\left[ 1 + 2(\as-1)
            \kappa(\btheta) \right]\;,
            \label{eq:103}
\end{align}
where in the second step we used the first-order Taylor expansion leading to Eq.\,(\ref{eq:numchange}), and in the last step we again made the weak lensing approximation $\mu\approx 1+2\kappa$.

The expectation value of the estimator (\ref{eq:ggl_gtx_estimator}) of the GGL signal is therefore affected by the local change of the source number density and becomes
\begin{align}
  \langle\hat\gt\rangle(\theta)
  &=\left\langle \kappa_\mathrm{g}(\btheta')\,\gt(\btheta'+\btheta;\btheta)\,{\frac{1}{\mu(\btheta'+\btheta;\btheta)}}\,{\frac{n_\mathrm{s0}[>s/\mu(\btheta'+\btheta;\btheta)]}
              {n_\mathrm{s0}(>s)}} \right\rangle\nonumber \\
&\approx\langle \kappa_\mathrm{g}(\btheta')\,\gt(\btheta'+\btheta;\btheta)\,\mu^{\as-1}(\btheta'+\btheta;\btheta) \rangle
  \label{eq:102}\\
&\approx
\langle \gt\rangle(\theta)
+2(\as-1) \langle \kappa_\mathrm{g}(\btheta')\,\gt(\btheta'+\btheta;\btheta)\,\kappa(\btheta'+\btheta;\btheta)\rangle\;. \nonumber
\end{align}
Thus, in the case of small magnifications, the bias is given by a third-order cross-correlation between the number density of foreground (lens) galaxies and the shear and convergence experienced by the background galaxies. This correlation is caused by the lensing effect of matter associated with the lens galaxies; hence, the bias is caused by magnification of sources by the matter at \zd. Equation\,(\ref{eq:102}) ignores the effect of intervening matter since this is sub-dominant for the source galaxy sample. Given that the bias term differs from the GGL signal by one order in the convergence, and that the characteristic convergence dispersion is of order $10^{-2}$, we expect that magnification of sources biases the GGL signal at the level of $\sim 1\%$.

Interestingly, the third-order correlator in the final expression of Eq.\,(\ref{eq:102}) is related to the galaxy-shear-shear correlator that was introduced by \citet{Schneider2005} as one of the $G_\pm$ quantities of galaxy-galaxy-galaxy lensing, since $\kappa$ and $\gamma$ are linearly related. Thus, from measurements of the galaxy-shear-shear correlations in a survey, this bias term can be directly estimated. We note that such measurements have already been successfully conducted \citep[e.g.][]{Simon2008,Simon2013}. A more quantitative description of this correction, which will be relevant for precision GGL studies in forthcoming surveys like \textit{Euclid} and LSST, is beyond the scope of this paper and will be done at a later stage.

An approximate, more intuitive way of describing the magnification of sources by lenses is provided by assuming that each lens is located at the centre of a halo of mass $m$.  In the case of no magnification, the expected tangential shear signal $\gt(\theta) = \ev{\est{\gamma}_{\mathrm{t}} (\theta)}$ can be expressed as\footnote{A more comprehensive halo model is discussed in Sect.\,\ref{sec:magnification_bias_in_halo_mass_estimates}.}
\begin{equation}
    \label{eq:ggl_expected_gt_without_magnification}
    \begin{split}
        \gt (\theta) &= 
        \int \dd \zs \, \pzs(\zs)
        \int \dd \zd \, \pzd(\zd)
        \int \dd m \, \pmforzd(m | \zd)
        \\&\quad\times
        \gt (\theta| m, \zd,\zs) \;
    \end{split}
\end{equation}
for a population of sources with redshift distribution $\pzs(\zs)$, a population of lenses with redshift distribution $\pzd(\zd)$, and a conditional distribution $\pmforzd(m| \zd)$ of the masses $m$ of the halos in which the lens galaxies reside. The mean tangential shear profile $\gt(\theta|m,\zd, \zs)$ for lenses with halo mass $m$ at redshift $\zd$ and sources at redshift $\zs$ can be factorised,
\eq{
    \label{eq:gammainf}
    \gt (\theta| m, \zd,\zs) = \gamma_\infty (\theta| m, \zd) \frac{\Dds}{\Ds} \;,
}
where $\Dds = D(\zd, \zs)$, and \(\gamma_\infty\) is the mean shear profile for (hypothetical) sources at infinite distance.

Equation \eqref{eq:ggl_expected_gt_without_magnification} assumes that the source number density is statistically independent of the lens positions. However, the observed number density of sources may change behind lenses due to magnification by the lenses. The expected magnification is a function of angular separation, the source and lens redshift, and the lens halo mass. For a flux-limited sample, the expected shear signal~\eqref{eq:ggl_expected_gt_without_magnification} then changes to:
\begin{equation}
    \label{eq:ggl_expected_gt_with_magnification_of_sources_by_lenses}
    \begin{split}
        \gt (\theta) &= 
        \Biggl[
        \int \dd \zs\, \pzs(\zs)
        \int \dd \zd\, \pzd(\zd)
        \int \dd m\, \pmforzd(m | \zd)
        \\&\quad\times
        \;\mu (\theta| m, \zd,\zs)^{\as(\zs) - 1}
        \biggr]^{-1} 
        \\&\quad \times
        \int \dd \zs\, \pzs(\zs)
        \int \dd \zd\, \pzd(\zd)
        \int \dd m\, \pmforzd(m | \zd)
        \\&\quad\times
        \;\mu (\theta| m, \zd,\zs)^{\as(\zs) - 1}\,
        \gt (\theta| m, \zd,\zs) \;,
    \end{split}
\end{equation}
where $\mu (\theta| m, \zd,\zs)$ denotes the mean magnification of sources at redshift $\zs$ and separation $\theta$ by lenses at redshift $\zd$ with halo mass $m$, and $\as(\zs)$ denotes the slope of the source counts at redshift $\zs$ at the source flux limit as in Eq.\,(\ref{eq:faintendslope}). We obtained the corresponding expression for a volume-limited source sample by replacing $\as$ by zero in Eq.\,\eqref{eq:ggl_expected_gt_with_magnification_of_sources_by_lenses}.

We may also assume that $\gt(\theta| m, \zd,\zs)$ and $\mu (\theta| m, \zd,\zs)$ are larger when the halo mass $m$ is larger (in the weak-lensing regime). Then, lens galaxies in more massive halos appear under represented in the estimator~\eqref{eq:ggl_gtx_estimator}, and the expected shear signal~\eqref{eq:ggl_expected_gt_with_magnification_of_sources_by_lenses} is lower than the prediction~\eqref{eq:ggl_expected_gt_without_magnification} ignoring magnification when $\as <1$. For $\as = 1$, the effect vanishes. For $\as > 1$, the number of source-lens pairs with more massive lenses in the GGL estimator~\eqref{eq:ggl_gtx_estimator} is enhanced more by magnification, and thus one expects a shear signal~\eqref{eq:ggl_expected_gt_with_magnification_of_sources_by_lenses} that is larger than the prediction~\eqref{eq:ggl_expected_gt_without_magnification}. When neglecting magnification, this may cause biases in the estimation of the mean halo mass.

As Eq.\,\eqref{eq:ggl_expected_gt_with_magnification_of_sources_by_lenses} indicates, magnification may also affect the observed redshift distributions of the lenses and sources. Furthermore, the magnification profile $\mu (\theta| m, \zd,\zs)$ usually has a strong radial dependence, being large for radii close to the Einstein radius, but rapidly dropping to values close to unity for larger radii. Thus, the magnification effects on the GGL signal are stronger for smaller radii. When neglecting magnification, this may cause additional biases when estimating parameters such as the halo concentration, and also when estimating the width of the halo-mass distribution.

\subsection{Mock data production}

We made use of ray-tracing results through the Millennium Simulation \citep[MS,][]{Springel2005,Hilbert2009}, from which we obtained tangential shear profiles using a fast-Fourier-transform (FFT) method. The method follows and improves the one presented in \citet{Unruh2019}. Specific details are given in Appendix \ref{sec:mockdata}.

For the radial binning of the tangential shear profile, we chose \(N_\mathrm{bin} = 16\) logarithmically spaced bins between \(\theta_\mathrm{in} = 0.6'\) and \(\theta_\mathrm{out} = 17.5'\). We estimated the error on the shear signal with a jackknife method that measures the field-to-field variance of the 64 fields. Then, we repeated the whole process while replacing the lens galaxies with random positions to obtain the shear estimate that is caused by the long modes in the matter density field, as well as boundary effects as recommended by \citet{Singh2017}. The random signal \(\est{\gamma}_\mathrm{rand}\) was subtracted from original signal \(\est{\gamma}_\mathrm{t} \to \est{\gamma}_\mathrm{t} -  \est{\gamma}_\mathrm{rand}\). For convenience, we dropped the hat to distinguish the estimator $\est{\gamma}_{\mathrm{t}}$ from theoretical expectations $\gt = \ev{\est{\gamma}_{\mathrm{t}}}$ in the following.

\begin{figure}[htbp]
    \centering
    \includegraphics[width=.49\textwidth]{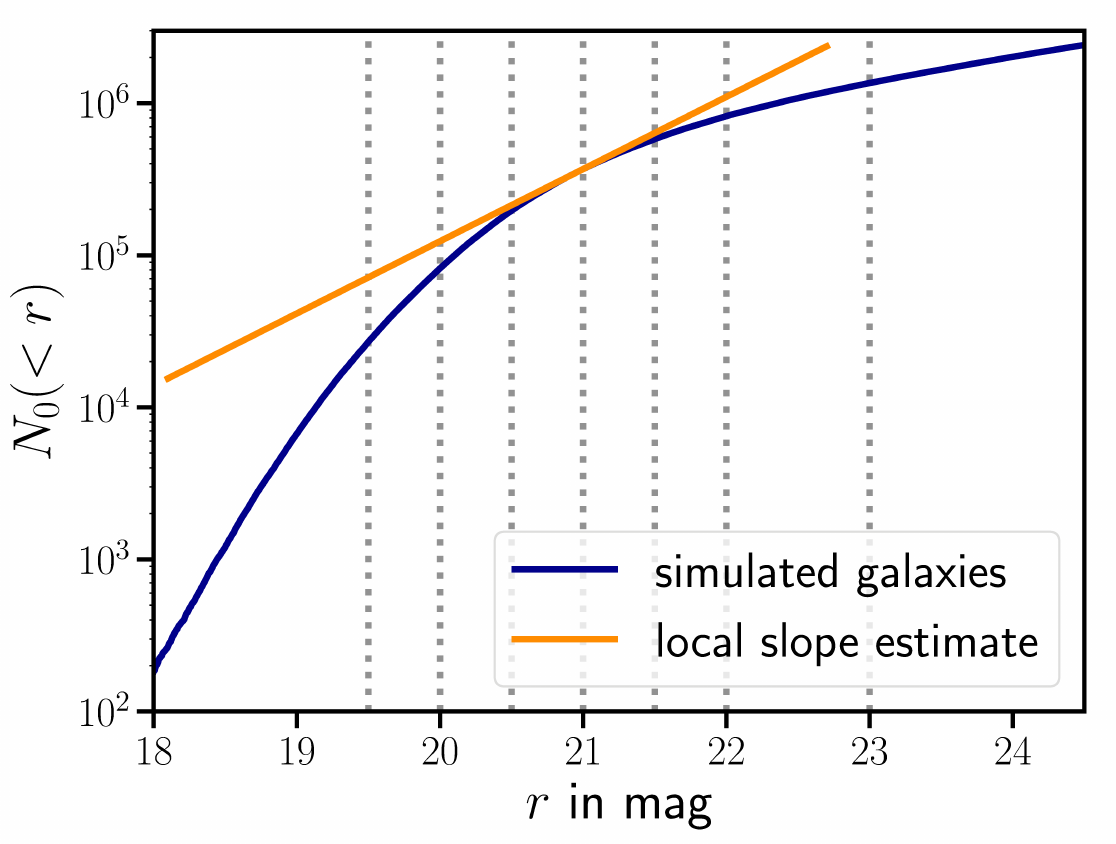}
    \caption{Cumulative number of galaxies \(N_0\) is shown as a function of \(r\)-band magnitude for a field of $64\times 16\,\degt^2$ at redshift \(z=0.41\). The dotted vertical lines indicate the magnitude cuts listed in Table~\ref{tab:magvsa}b. Finally, the tangential curve shows the local slope \(\alpha\) at \(r=21\,\mathrm{mag}\).}
    \label{pic:numcounts}
\end{figure}

For our analyses, we wanted to obtain samples of galaxies with different local slopes \(\alpha\) of the source counts, which depend on redshift and limiting magnitude. For a given lens galaxy sample with a magnitude cut corresponding to a flux limit \(s_\mathrm{lim}\), we estimated the local slope (\ref{eq:faintendslope}) by finite differencing around \(s_\mathrm{lim}\). Figure\,\ref{pic:numcounts} illustrates the cumulative number of galaxies at redshift \(\zd=0.41\) for the whole simulated field of $64\times 16\,\degt^2$. Several magnitude cuts are indicated, as well as one example of a tangential curve at \(s_\mathrm{lim}=21\,\mathrm{mag}\) with slope \(\alpha = 1.06\). For all GGL measurements in this work, we fixed the source redshift to \(\zs=0.99\). To find the local slopes \as\ of the source galaxy counts, we varied the limiting magnitude in the \(r\)-band filter (see Table~\ref{tab:magvsa}a).  For the local slope \ad\ of the lens galaxy counts, we further varied the lens redshift \zd,\ as well as the limiting magnitude, as shown in Table~\ref{tab:magvsa}b and \ref{tab:magvsa}c.
\begin{table}[htbp]
    \centering
    \caption{The local slopes for the source galaxy counts \as\ and the lens galaxy counts \ad\ are given as a function of redshift and as a function of the limiting flux. Either redshift or limiting flux is always kept fixed, while the other quantity is varied, as shown below. The limiting flux is given in terms of $r$-band magnitude.}
    \label{tab:magvsa}
    \begin{tabular}{cc|cc|cc}
        \multicolumn{2}{c|}{(a)} & \multicolumn{2}{c|}{(b)} & \multicolumn{2}{c}{(c)} \\
        \multicolumn{2}{c|}{fix \(\zs=0.99\)} & \multicolumn{2}{c|}{fix \(\zd=0.41\)} & \multicolumn{2}{c}{fix \(\slimd=22\)} \\
        &&&&&\\ 
        \slims & \as & \slimd & \ad & \zd & \ad \\
        \hline\hline
        22.0 & 2.89 & 19.5 & 2.71 & 0.24 & 0.46 \\
        22.5 & 2.33 & 20.0 & 2.03 & 0.41 & 0.66 \\
        23.0 & 1.75 & 20.5 & 1.56 & 0.51 & 0.98 \\
        23.5 & 1.29 & 21.0 & 1.06 & 0.62 & 1.58 \\
        24.0 & 0.93 & 21.5 & 0.85 & 0.83 & 2.49 \\
        24.5 & 0.68 & 22.0 & 0.66 & & \\
        25.0 & 0.58 & 23.0 & 0.44 & & \\
        26.0 & 0.45 & &  & &
    \end{tabular}
\end{table}

\section{Magnification effects on background galaxies}
\label{sec:magnification_effects_on_background_galaxies}

Lens and source galaxies are affected by magnification. To understand this impact in more detail, we first discuss how the shear profile changes when only source galaxies are magnified. In the following, we describe how we generated an arbitrary number of mock source catalogues, with and without magnification bias included. Results from the Millennium Simulations are then presented, and the magnification-induced bias in the GGL signal is compared to the prediction of the analytical model presented in Sect.\,\ref{Sect:Msl}.

\subsection{Magnification switched off}

To switch magnification off, we simply chose random source positions. We set the number of galaxies to \(N_\mathrm{s,0}=10^7\) per $16\,\degt^2$ field to keep the impact of noise low.

\subsection{Magnification switched on}

Magnification changes the number counts of observed galaxies on the sky. Using the ray-tracing data, we obtained the cumulative number counts of the galaxies as a function of magnification-corrected flux, \(n_\mathrm{s,0} (>s_0)\). We obtained the local expected number counts of galaxies by adjusting the flux limit to \(\slims/\mu(\btheta)\) at each position \btheta\ and using the first equality in Eq.\,(\ref{eq:103}).
%
%
We further scaled the number counts so that for \(\mu=1\) the expected number of source galaxies is \(N_\mathrm{s}=10^7\) per field of solid angle \(A=4^\circ\times4^\circ\). The threshold of finding a source at a grid position~\btheta\ is then \(T(\btheta) = n_\mathrm{s}(>\slims; \btheta) \, A / N_\mathrm{pix}\), where we restricted \(T(\btheta)\) to be smaller than unity. Finally, we drew a uniform random number \(P(\btheta)\) between zero and one for each position. A source galaxy was placed at a position \btheta\ if \(T(\btheta) > P(\btheta)\). The `magnification off' method can be recovered if we insert \(\mu =1\) for all \btheta\ in Eq.\,(\ref{eq:103}).

\subsection{Results}
\label{ssec:sourcebias}
\begin{figure*}[htbp]
    \centering
    \includegraphics[width=\textwidth]{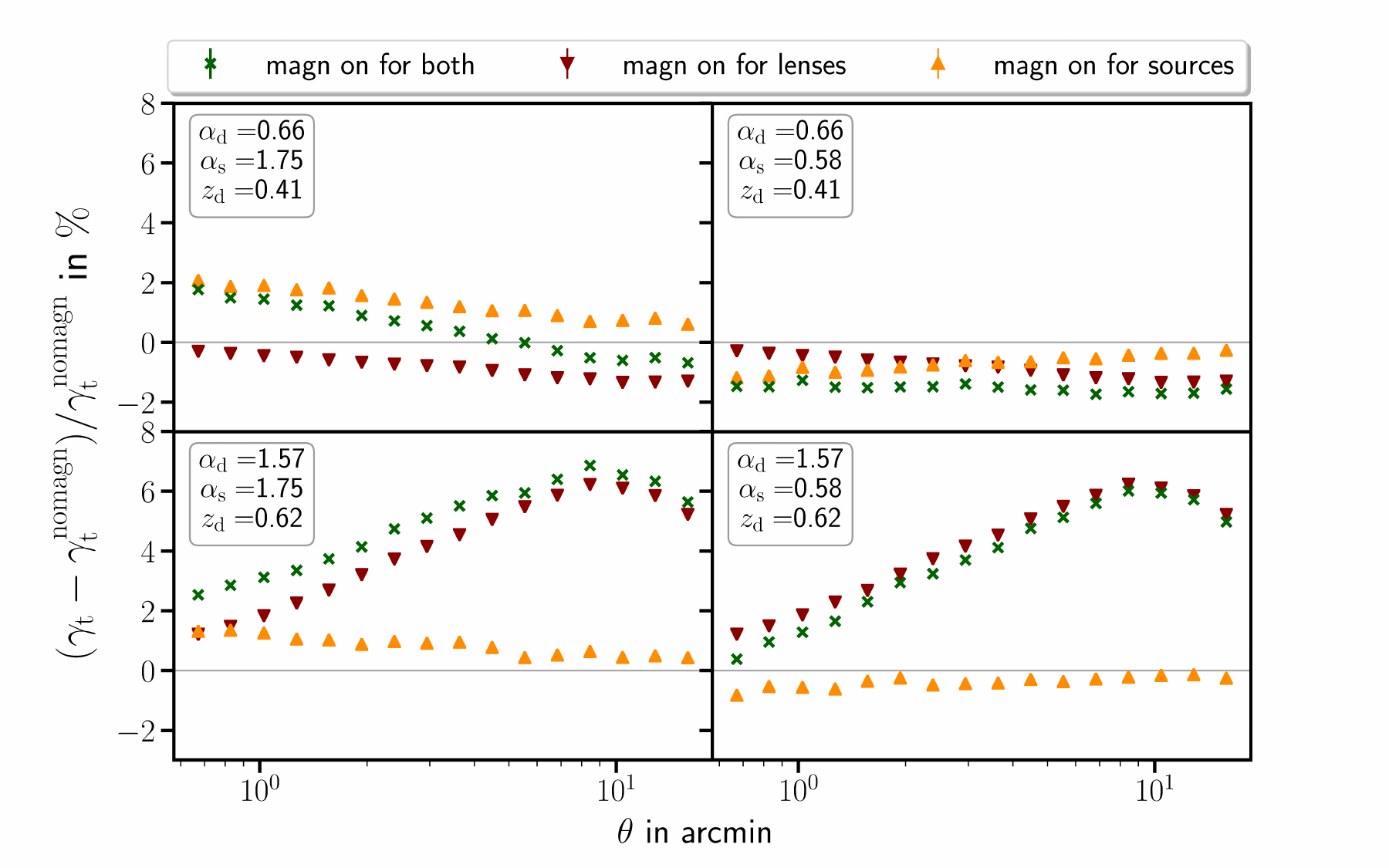}
    \caption{Relative difference between shear profiles with and without magnification. The redshifts for the lenses are \(\zd=0.41\) in the upper panels and \(\zd=0.62\) in the lower panels; the source redshift is kept fixed at \(\zs=0.99\). The limiting magnitude of the lenses is \(22\,\mathrm{mag}\) in the \(r\)-band, and for the sources it is \(23\) and \(25\,\mathrm{mag}\) for \(\as=1.75\) and \(0.58\), respectively. The red `downward' triangles indicate shear profiles that only have magnification in the lens galaxy population, while the orange `upward' triangles show the influence of magnification for source galaxies only. It can be seen that a local slope $>1$ of lens or source population leads to an enhanced signal, whereas \(\ads<1\) causes a reduced signal. The green crosses display a measurement closest to real observations, i.e. where magnification affects both source and lens galaxy populations. A reduction or enhancement depends on both slopes \ads, as well as the redshifts of lenses and sources \zds. In all cases, the shape of the shear profile changes.}
    \label{pic:shearprofs}
\end{figure*}

The relative impact of magnification of sources on a tangential shear profile is shown by the orange `upward' triangles in Fig.\,\ref{pic:shearprofs}.
As expected, the net effect depends on the local slope \as; the effect is typically of the order of \(1\) to \(2\%\) per bin. In the two panels on the left-hand side in Fig.\,\ref{pic:shearprofs}, the local slope \as\ is larger than unity, and the shear signal is enhanced; while in the two panels on the right, \(\as <1\), which reverses the effect. Also, the magnification effect is stronger for smaller separations $\theta$.
\begin{figure*}[htbp]
    \centering
    \sidecaption
    \includegraphics[width=.617\textwidth]{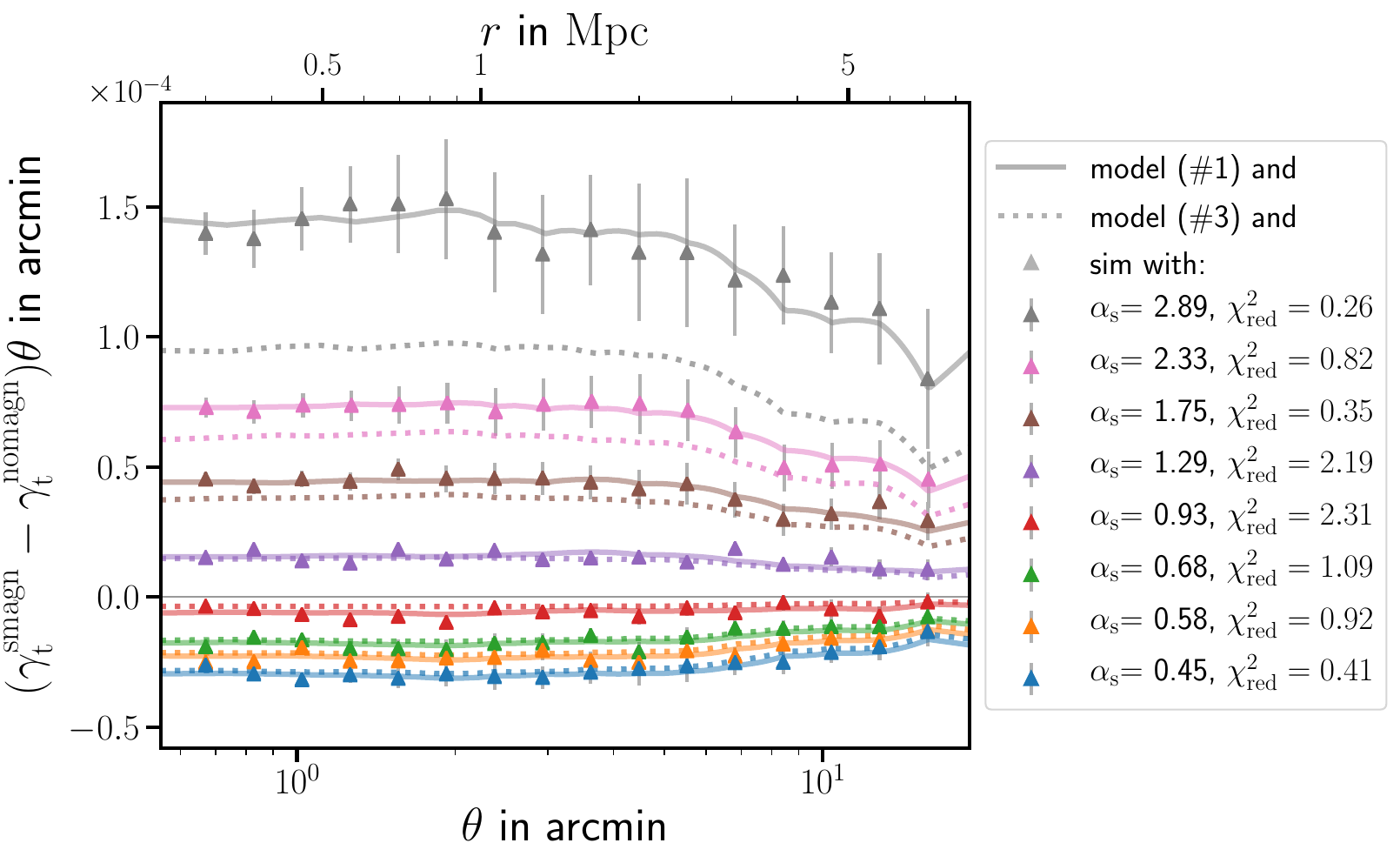}
    \caption{Absolute difference between shear profiles with and without magnification of sources, for different local slopes \as\ shown in the legend. The solid lines correspond to the first expression in Eq.\,(\ref{eq:102}) and dotted lines correspond to its approximation in the third line. The upper scale shows comoving transverse separation, and the redshifts are \(\zs=0.99\) for the sources and \(\zd=0.41\) for the lenses. The brown triangles with \(\as = 1.75\) and the orange ones \(\as=0.58\) are directly comparable to the orange triangles in the upper panels of Fig.\,\ref{pic:shearprofs}. We also show the goodness-of-fit parameter \(\chi^2_\mathrm{red}\) with 16 degrees of freedom for the solid line.}
    \label{pic:sheardiff_source}
\end{figure*}
This is seen more clearly in Fig.\,\ref{pic:sheardiff_source}, where the absolute difference of source-magnified to magnification-corrected shear profiles is compared. The shear profiles vary with \as\ for constant redshifts \zds\ according to Table~\ref{tab:magvsa}a. The difference between the expected and `measured' shear profiles in Fig.\,\ref{pic:sheardiff_source} is the bias that we estimated in Sect.\,\ref{Sect:Msl} and gave in Eq.\,(\ref{eq:102}). We calculated the first and the third line of~(\ref{eq:102}) using the numerical data and show them as solid and dotted lines, respectively. Both models are in good agreement with the numerical data for moderate \as. However, for very steep \(\as\gtrsim2,\) the weak lensing approximation \(|\mu-1|\ll1\) is not sufficient anymore; although large magnifications are rare, they affect the number counts significantly.

We define the mean fractional difference between a shear profile with and without magnification for all bins as
\eq{
    \label{eq:sheardiff}
    \tgt = \frac{1}{N_\mathrm{bin}}\,\sum_{i=1}^{N_\mathrm{bin}} \frac{{\gt}_i - {\gt}^\mathrm{nomagn}_i}{{\gt}^\mathrm{nomagn}_i} \;,
}
where we stress that a difference of \(\tgt =0\) is not necessarily equivalent to an unaltered shear profile. However, we only applied this estimator to the orange `upward' and red `downward' triangles seen in Fig.\,\ref{pic:shearprofs}, which display either a positive or a negative sign for all angular scales investigated.

\begin{figure}[htbp]
    \centering
    \includegraphics[width=.49\textwidth]{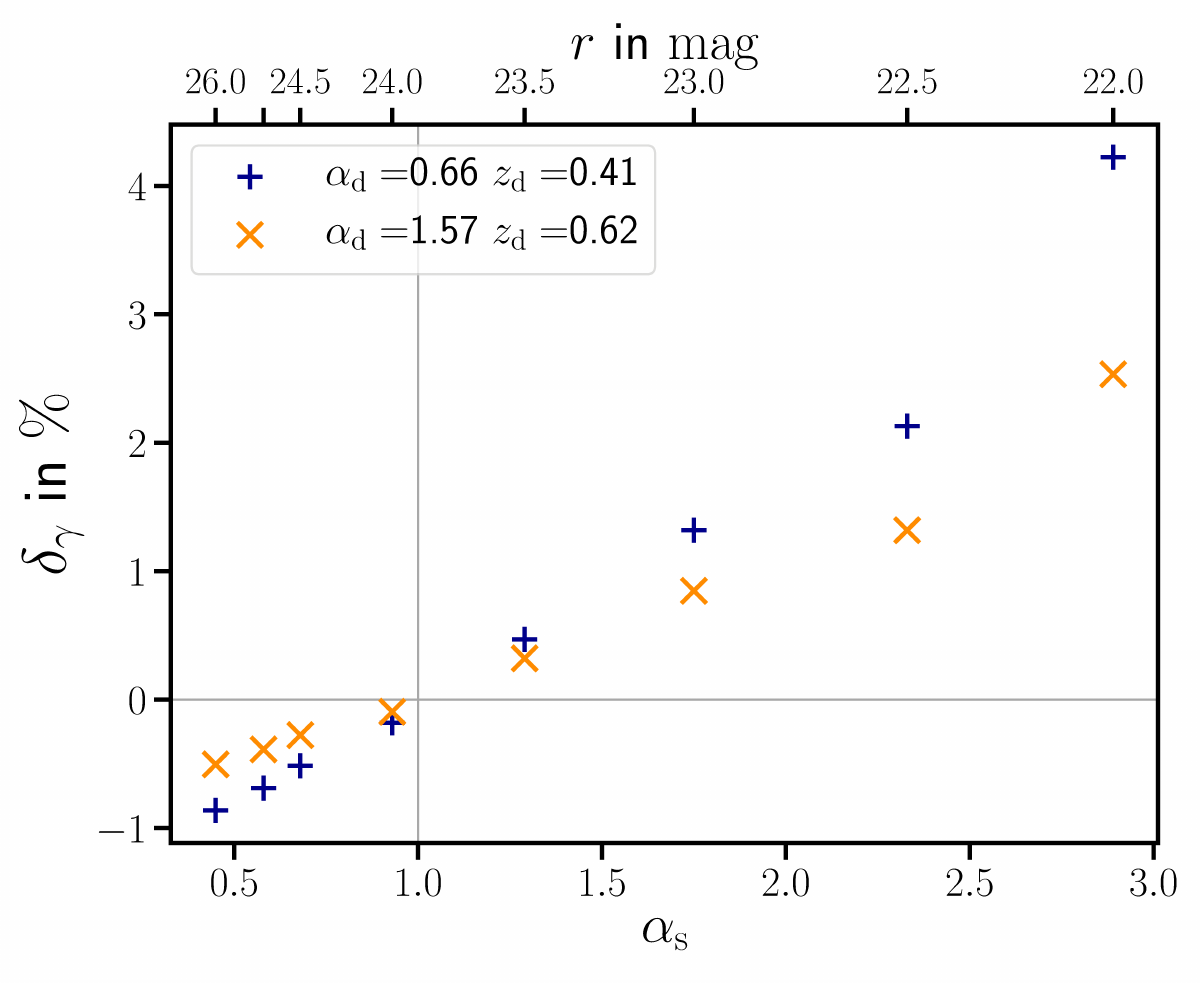}
    \caption{Fractional change of the shear profile by magnification is displayed, where the magnification is only turned on for the source galaxies at \(\zs=0.99\) (Table~\ref{tab:magvsa}a). At a local slope of \(\as \approx 1\) the impact of magnification flips its sign. The relation between \as\ and \tgt\ is almost linear. Different redshifts of the lenses \(\zd=0.41\) (plus) and 0.62 (cross) affect the impact of the sources' magnification; the  \(r\)-band limiting magnitude of lenses is fixed to \(22\,\mathrm{mag}\).}
    \label{pic:evolsource}
\end{figure}

Results for \tgt\ as a function of \as\ for constant \zs\ can be seen in Fig.\,\ref{pic:evolsource}. We selected the source galaxies according to Table~\ref{tab:magvsa}a, for which \tgt\ is almost linear in \as, as expected from Eq.\,(\ref{eq:102}) in the weak lensing approximation, although the slope of the linear relation depends on different \zd\ (Table~\ref{tab:magvsa}c). The maximum shear difference of \(4\%\) is found for the largest \as.

\section{Magnification effects on foreground galaxies}
\label{sec:magnification_effects_on_foreground_galaxies}

In this section, we investigate the influence of magnification on lens galaxies. We follow the structure from the previous section, meaning we obtain and analyse results from the ray-tracing data and then compare those to the analytic estimate presented in Sect.\,\ref{Sect:MlLSS}.

\subsection{Magnification switched on}

The galaxies in the Henriques catalogue are affected by magnification by design. Hence, to create a catalogue including magnification, we simply extracted lens positions from galaxies brighter than a magnitude limit \(\slimd\) and assigned them to their nearest grid point.

\subsection{Magnification switched off}
To switch off magnification in the mock data, we undid the magnification as follows. As was done for the sources, the magnification-corrected flux \(s_0\) can be easily recovered from the magnification given in the ray-tracing catalogue and the apparent flux of lens galaxies. The galaxy's apparent position, however, is shifted on the sky compared to its unlensed position. Unfortunately, the unlensed position cannot be recovered from the simulated data, and the absolute amount of shifting is of the order of an arcminute and depends on redshift \citep{Chang2014}. Hence, we must aim to create a sample of galaxies that is not affected by magnification in a different way.

As was outlined in Sect.\,\ref{ssec:magneffects}, for both local slopes \mbox{\(\ads =1 \)}, the magnification leaves the observed number counts unaffected, and thus the shear profile \gt\ unbiased. Therefore, we transformed the magnification-corrected flux distribution such that the galaxy counts obey \(n'_0 \propto s_0' \,\!^{-1}\). As a reference point, we chose the number of galaxies at limiting magnitude \mbox{\(n_0(>\slimd) = n_0' (>\slimd')\)}. This results in the following mapping from observed magnification-corrected flux, \(s_0\), to the transformed flux:
\eq{
    s_0'(s_0) = \slimd \, \frac{n_0(\slimd)}{n_0(s_0)} \;.
}
In other words, this new flux scale distorts the number counts of lenses in such a way that in the transformed flux system, $\alpha'=1$, the lens galaxy counts are unaffected by magnification bias (although the individual lens positions are not). We can now calculate the observed transformed number density \(n'(>s')\), where \(s' = \mu \,s_0'\). To create a lens galaxy sample free from magnification, we again chose only those galaxies that are brighter than the given flux limit \(s'>\slimd\). This leads to a different selection of galaxies for the original and the transformed number density.

We tested this approach with two consistency checks. The first is based on the fact that in the method described above, the total number of galaxies has to be conserved. This is true for all the lens redshifts used. For a magnitude cut of \(22\,\mathrm{mag}\) in the \(r\)-band, a lens redshift of \(\zd = 0.41,\) and the full field of view of \(64 \times 16 \,\mathrm{deg}^2\), \(595\,348\) lenses are found with a cut in observed magnitude, and \(595\,355\) lenses are found with observed transformed magnitude. Compared to the original fluxes, a detailed analysis showed that in the transformed flux system, \(917\) galaxies became brighter than \(22\,\mathrm{mag,}\) while \(924\) galaxies became dimmer, leaving the overall number count almost unchanged. The tiny difference in seven galaxies is due to the fact that the number count function is discretely sampled.

The second consistency check uses a null test: the so-called shear-ratio test \citep[SRT,][]{jain2003}, which is based on Eq.\,(\ref{eq:gammainf}), 
\eq{
    \label{eq:SRT}
    T(\theta; \zd, z_{\mathrm{s}_1}, z_{\mathrm{s}_2}) := \frac{\gt(\theta; \zd, z_{\mathrm{s}_1})}{\gt(\theta; \zd, z_{\mathrm{s}_2})} - \frac{D_{\mathrm{ds}_1}}{D_{\mathrm{s}_1}}\frac{D_{\mathrm{s}_2}}{D_{\mathrm{ds}_2}} \;,
}
for which we expect \(T(\zd, z_{\mathrm{s}_1}, z_{\mathrm{s}_2})=0\) for two source populations at redshifts \(z_{\mathrm{s}_1}\) and \(z_{\mathrm{s}_2}\) in the absence of magnification effects. For the test, the location of the same lens galaxies and the shear from two source galaxy populations at different distances \(D_{\mathrm{s}_{1,2}}\) were used. A ratio of the tangential shear estimates is equal to the ratio of the corresponding angular-diameter distances, while the lens properties drop out. As shown in \citet{Unruh2019}, the SRT is strongly affected by lens magnification. The impact is stronger for higher lens redshifts and smaller line-of-sight separation of lenses and sources. Therefore, we performed the SRT for lenses selected with and without magnification for two different lens redshifts. We performed the SRT by taking a weighted integral of \( T(\theta; \zd, z_{\mathrm{s}_1}, z_{\mathrm{s}_2})\) over \(\theta\) from \(\theta_\mathrm{in}\) to \(\theta_\mathrm{out}\) as in \citet{Unruh2019}. In the case that includes magnification, we recovered the results from \citet{Unruh2019}. For lenses that are selected with corrected magnification, the SRT performs better by a factor of \(\gtrsim100\). We give results for two example redshift combinations in Table~\ref{tab:SRTres}. The corrected SRT still shows a slight scatter due to the statistical noise in the data, coming from the lensing by the large-scale structure in each of the 64 fields, and a bias that arises from shifting lens galaxies to their nearest grid point.
\begin{table}[htbp]
    \centering
    \caption{Shear-ratio test (\ref{eq:SRT}) performed for two example cases to demonstrate the removal of the magnification from the lens galaxies.}
    \label{tab:SRTres}
    \begin{tabular}{p{5mm}p{5mm}p{5mm}|cr}
        \multicolumn{3}{c}{redshifts} & \multicolumn{2}{c}{SRT result \(T(\zd, z_{\mathrm{s}_1}, z_{\mathrm{s}_2})\)} \\
        \zd & \({\zs}_{1}\) & \({\zs}_{2}\) & magn & \multicolumn{1}{c}{no magn}  \\
        \hline\hline
        0.41 & 0.46 & 0.51 & \((5.6\pm1.4)\times10^{-2}\) & \((7.0\pm13.7)\times10^{-4}\) \\
        0.83 & 0.91 & 0.99 & \((1.3\pm0.2)\times10^{-1}\) & \((-5.1\pm10.5)\times10^{-4}\)
    \end{tabular}
\end{table}
\subsection{Results}
The red `downward' triangles in Fig.\,\ref{pic:shearprofs} show the relative impact of magnification on \gt. Similar to the results given in Sect.\,\ref{ssec:sourcebias}, where the magnification of the source galaxies is discussed, the local slope \ad\ determines whether the shear signal is enhanced or reduced. The upper panels of Fig.\,\ref{pic:shearprofs} show that the shear profiles are reduced for a local slope \ad\ that is smaller than unity at redshift \(\zd=0.42\), while the lower panels display shear profiles with \(\ad >1\) at a higher redshift \(\zd=0.62,\) where the reverse effect is observed. The panels with higher \zd\ show larger magnification effects; relative deviations by up to \(7\%\) in a single bin can be seen. In general, the shear signal is more strongly affected at larger separations $\theta$ from the lens centre until it reaches a maximum at \(\approx 8'\) for \(\zd=0.62\) and \(\approx 10'\) for \(\zd=0.41\); for even larger separations, the magnification effect becomes relatively weaker.

\begin{figure*}[htbp]
    \centering
    \sidecaption
    \includegraphics[width=.7\textwidth]{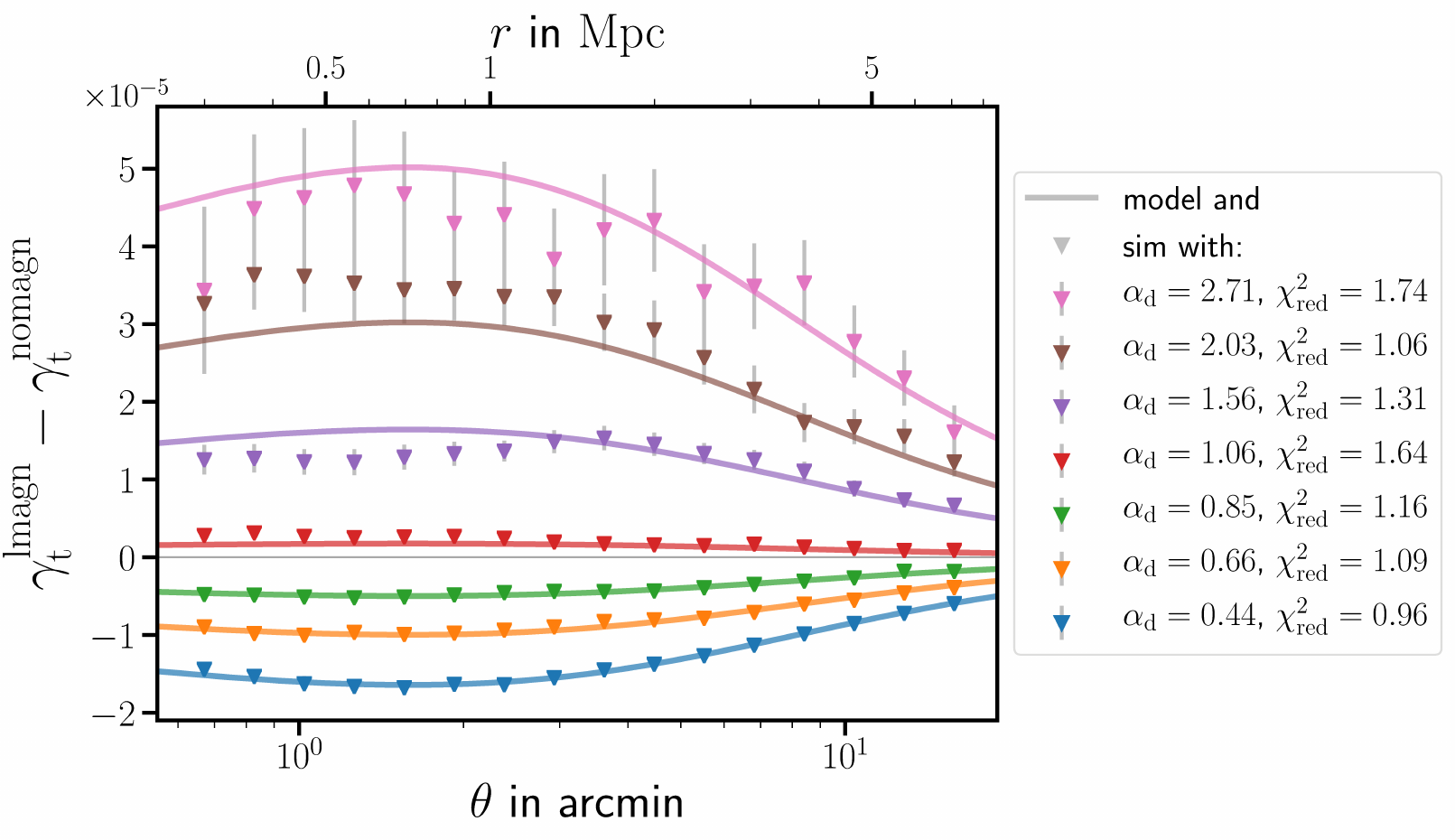}
    \caption{Absolute difference between shear profiles with and without magnification of lenses. The upper scale shows comoving transverse separation and the shear difference is shown for several limiting magnitudes of the lens galaxies, which yield the local slopes \ad\ shown in the legend. Also, the goodness-of-fit parameter \(\chi^2_\mathrm{red}\) with 16 degrees of freedom is indicated in the legend. The redshifts are \(\zs=0.99\) for the sources and \mbox{\(\zd=0.41\)} for the lenses.}
    \label{pic:sheardiff_lens}
\end{figure*}
A comparison of numerical results to the analytic estimate (\ref{eq:lensmagn}) can be seen in Fig.\,\ref{pic:sheardiff_lens}. The absolute difference between shear profiles affected by a magnification of lens galaxies and those unaffected by magnification is plotted. The triangles show numerical results, while lines indicate our analytical model for \(2\left[\ad(\zd) - 1 \right]\,\gt^\mathrm{LSS}\). We employed the reduced \(\chi^2\)-test as an estimator for the goodness of our model and find that all models are in good agreement with the data for the considered angular scales. However, the local slope is not necessarily a sufficiently good quantity for the analytic correction if the local slope \ad\ becomes very steep, meaning when the luminosity function is not well approximated by a power law anymore (cf.~Fig.\,\ref{pic:numcounts}). However, the analytic correction still reduces the impact of magnification significantly.

To explore the dependencies of the mean fractional shear difference \tgt\ (Eq.\,\ref{eq:sheardiff}) on \ad, we altered the lens properties according to Table~\ref{tab:magvsa}b. 
\begin{figure}[htbp]
    \centering
    \includegraphics[width=.49\textwidth]{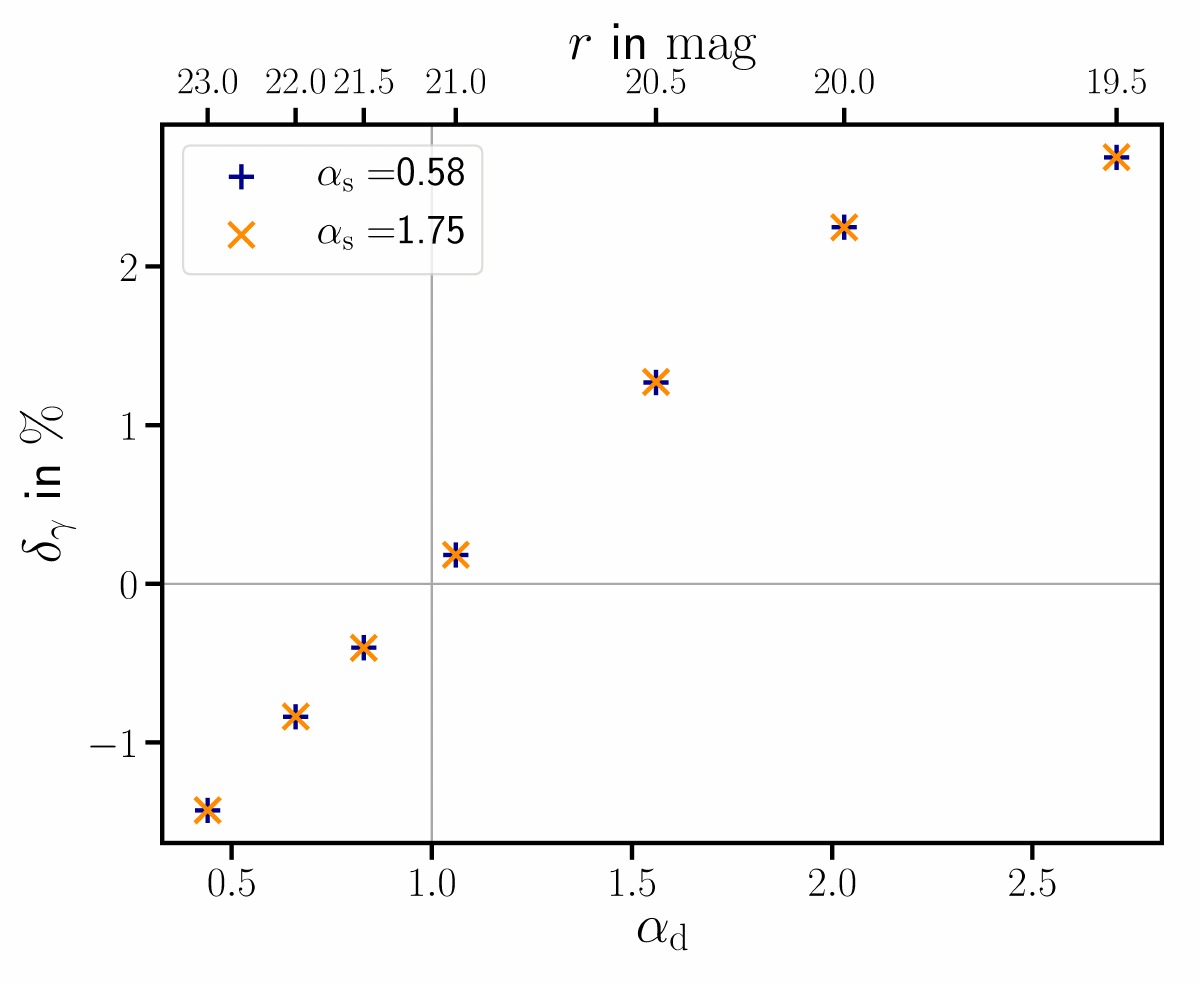}
    \caption{To change the lens properties according to Table~\ref{tab:magvsa}b, we show the behaviour of mean fractional shear difference \tgt\ as a function of local slope for lens galaxies \ad\ and \(r\)-band-limiting magnitude. The source and lens redshift is the same for both cases, i.e., \(\zd=0.41\) and \(\zs=0.99\). The effect is independent of the sources local slope \as.}
    \label{pic:evollens2}
\end{figure}
Results can be seen in Fig.\,\ref{pic:evollens2} and show the impact of magnification for constant lens redshift \zd\ and varying limiting magnitude, for two different local slopes \as\ of the sources. The impact on the shear profile is almost exactly the same in both local slopes, with a small residual noise that is present in the data. The dependence of \tgt\ on \ad\ is similar to the one in Fig.\,\ref{pic:evolsource}, which shows the magnification for source galaxies only. The shear profile changes by up to \(3\%\) in the mean.

\begin{figure}[htbp]
    \centering
    \includegraphics[width=.49\textwidth]{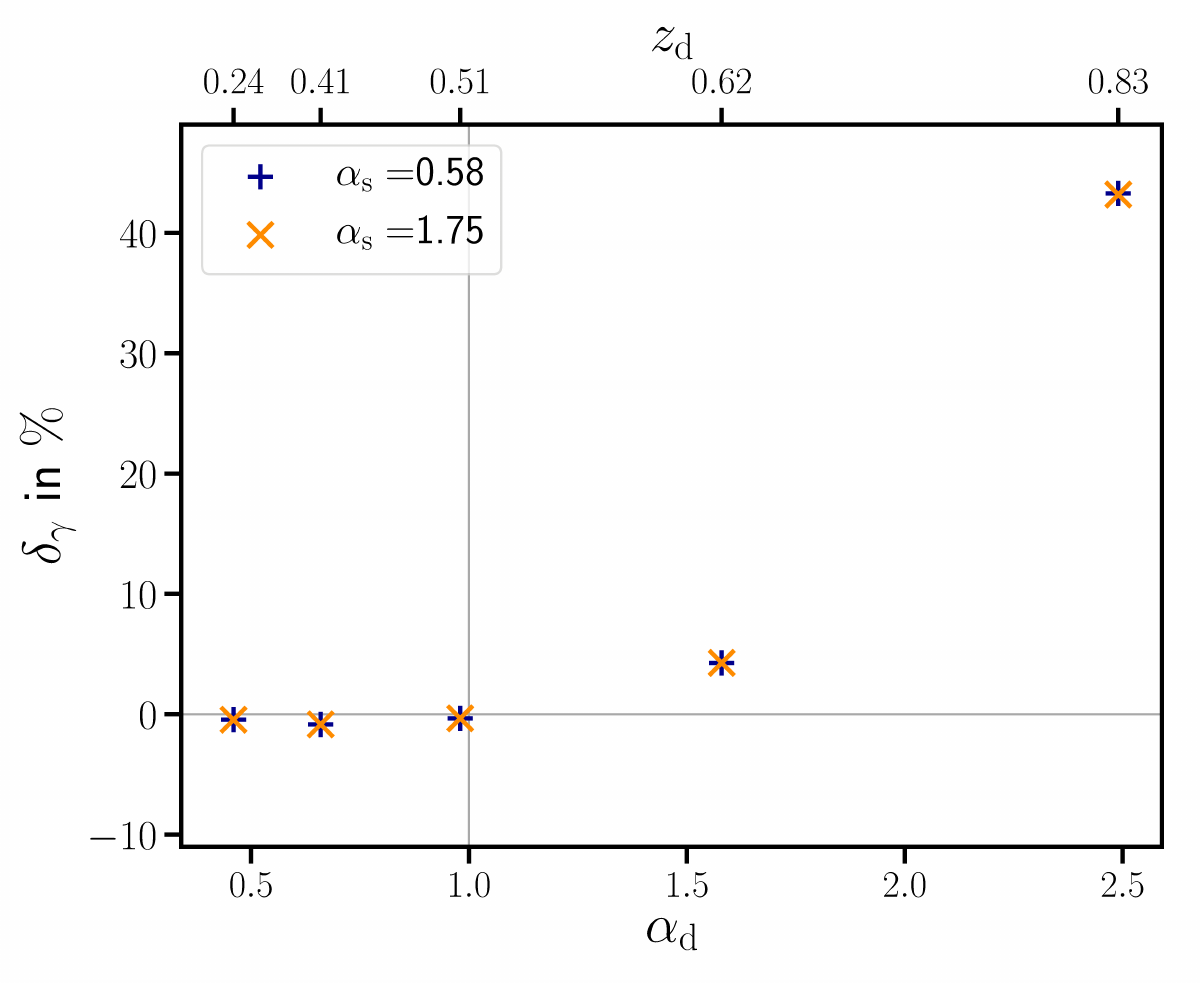}
    \caption{For a magnification of lens galaxies only, the impact on the shear profile is shown as a function of \ad\ and \zd\ (Table~\ref{tab:magvsa}c) for fixed flux limit of \(r=22\,\mathrm{mag}\). The redshift \zd\ is given on the (non-linear) top axis. Again, the effect switches its sign at \(\ad \approx 1\). It then rises up to \(45\%\) for larger \ad, while less impact is seen for \(\ad<1\). Different local slopes of the sources \as\ at \(\zs=0.99\) leave the lens' magnification unaffected.}
    \label{pic:evollens}
\end{figure}
We further investigated the dependence of the magnification on the values of \ad\ and \zd\ for fixed \zs\ in Fig.\,\ref{pic:evollens}. \tgt\ is calculated for various lens redshifts with constant limiting magnitude for the lenses (see Table~\ref{tab:magvsa}c) and the same two values of \as\ as before. The figure shows that the sources' local slope does not affect the magnification induced by lens galaxies. The signal is again moderately reduced by \(1\%\) for \(\ad<1\) but is not monotonic anymore. For \(\ad>1,\) the relation deviates significantly from a linear one. Magnification is stronger for larger \ad\ and larger \zd, leading to deviations of up to \(45\%\) in the most extreme case.

Lastly, we combined the numerical methods and results that were addressed individually in Sects.\,\ref{sec:magnification_effects_on_background_galaxies} and \ref{sec:magnification_effects_on_foreground_galaxies}. The green crosses in Fig.\,\ref{pic:shearprofs} show numerical results from source and lens galaxy populations that are both affected by magnification, meaning the case of relevance for observational studies of GGL. The fractional change is approximately the sum of the bias of lens galaxies only plus the bias of source galaxies only. Hence, to first order, it is determined by \ad\ and \as. The sign of \tgt\ cannot easily be predicted if \(\ad<1\) and \(\as>1\), and vice versa. For \tgt\ the sign also depends on $\theta$ since the change of shear signal per bin behaves differently for magnified lenses and magnified sources.
\section{Magnification bias in halo-mass estimates}
\label{sec:magnification_bias_in_halo_mass_estimates}
\subsection{Estimating the mean halo mass of lenses}
The GGL signal is sensitive to the surface-mass density around lenses that differs from the average projected cosmological matter density. To infer from this the mean mass of a parent halo that hosts a typical lens galaxy, we used a halo-model prescription to describe the relation between galaxies and matter \citep{2002PhR...372....1C}. In this prescription, we expand the galaxy-matter power spectrum at redshift \zd\ of lenses and at comoving wave number $k$,
\begin{equation}
    P_\mathrm{gm}(k)= P_\mathrm{gm}^\mathrm{1h}(k)+P_\mathrm{gm}^\mathrm{2h}(k)\;,
\end{equation}
in terms of a one-halo term,
\begin{multline}
    P_\mathrm{gm}^\mathrm{1h}(k)=\\
    \int_0^\infty\frac{\mathrm{d} m\,n(m)\,m}{\Omega_\mathrm{m}\,\rho_\mathrm{crit}\,\bar{n}_\mathrm{d}}\;
    \tilde{u}_\mathrm{m}(k,m)\,
    \Big( \ev{N_\mathrm{cen}|m}+\ev{N_\mathrm{sat}|m}\,\tilde{u}_\mathrm{m}(k,m) \Big)\;,
\end{multline}
and a two-halo term,
\begin{multline}
    P_\mathrm{gm}^\mathrm{2h}(k)=\\
    \int_0^\infty\frac{\mathrm{d} m\,n(m)\,m\,b_\mathrm{h}(m)}{\Omega_\mathrm{m}\,\rho_\mathrm{crit}}\;
    \tilde{u}_\mathrm{m}(k,m)\,P_\mathrm{lin}(k)\,\\
    \times\int_0^\infty\frac{\mathrm{d} m\,n(m)\,b_\mathrm{h}(m)}{\bar{n}_\mathrm{d}}\;  
    \Big( \ev{N_\mathrm{cen}|m}+\ev{N_\mathrm{sat}|m}\,\tilde{u}_\mathrm{m}(k,m) \Big)\;.
\end{multline}
In this model, $\tilde{u}_\mathrm{m}(k,m)$ denotes the Fourier transform of a Navarro-Frenk-White \citep[NFW;][]{1996ApJ...462..563N} density profile for a virial halo mass $m$, truncated at the virial radius and normalised to $\tilde{u}_\mathrm{m}(k,m)=1$ for $k=0$ \citep{2001ApJ...546...20S} for the mass-concentration relation in \cite{2001MNRAS.321..559B}. We further denote the mean comoving number density of halos in the mass interval \mbox{$m_1\le m<m_2$} by $\int_{m_1}^{m_2}\mathrm{d} m\,n(m)$ \citep{1999MNRAS.308..119S}, the bias factor of halos of mass $m$ by $b_\mathrm{h}(m)$ \citep{2001ApJ...546...20S}, and the linear matter power spectrum by $P_\mathrm{lin}(k)$ \citep{1998ApJ...496..605E}. Finally, the mean number density of lenses is
\begin{equation}
    \bar{n}_\mathrm{d}=
    \int_0^\infty\mathrm{d} m\,n(m)\,
    \Big( \ev{N_\mathrm{cen}|m}+\ev{N_\mathrm{sat}|m} \Big) \;.
\end{equation}
This version of the halo model assumes a central galaxy at the centre of a halo whenever there are lens galaxies inside the halo, and satellite galaxies with a number density profile equal to the NFW matter density. For the mean number of central and satellite galaxies for a halo mass $m,$ we follow \citet{Clampitt2017} but with central-galaxy fraction $f_\mathrm{cen}\equiv1$,
\begin{align}
    \ev{N_\mathrm{cen}|m}&=
    \frac{1}{2}
    \left[ 1+\mathrm{erf}{\left(\frac{\log_{10}{(m/m_\mathrm{th})}}{\sigma_{\log m}}\right)} \right]\;;\\
    \ev{N_\mathrm{sat}|m}&=
    \ev{N_\mathrm{cen}|m}\,\left(\frac{m}{m_1}\right)^\beta\;,                     
\end{align}
where $\bol{\Theta}=(m_1, m_\mathrm{th}, \sigma_{\log m}, \beta)$ are four model parameters that determine the halo-occupation distribution (HOD) of our lenses, and $\mathrm{erf}(x)=2 \pi^{-1/2}\int_0^x\mathrm{d} t\,\mathrm{e}^{-t^2}$ is the error function. The model parameters have the following meaning: $m_\mathrm{th}$ determines at which mass scale $\ev{N_\mathrm{cen}|m}=1/2$; at halo mass $m_1$ the mean number of satellites equals that of central galaxies; $\sigma_{\log m}$ is the width of the HOD of centrals; and $\beta$ is the slope of the satellite HOD.

The matter-galaxy cross-power spectrum is related to the mean tangential shear by a Limber projection, which for lenses at \zd\ and sources at \zs\ is
\eq{
    \gt(\theta)=
    \frac{3H_0^2\,\Omega_\mathrm{m}}{2c^2}\; \frac{\Dds}{\Dd\,\Ds}\,
    \int_0^\infty\frac{\mathrm{d}\ell\,\ell}{2\pi}\;\mathrm{J}_2(\ell\theta)\, P_\mathrm{gm}\left(\frac{\ell+1/2}{(1+\zd)\,\Dd};\zd\right)\;.
}

We employed a maximum-likelihood estimator (MLE) to infer the mean halo mass of lenses from GGL. For this, we set \mbox{$\{\gamma_{\mathrm{t},i}(\bol{\Theta}) \,|\, i=1\ldots N_\mathrm{bin}\}$} as a set of $N_\mathrm{bin}$ measurements of the mean tangential shear in our mock data, obtained for different lens-source angular separation bins \(i\); the error covariance estimated from the measurements for bin $\theta_i$ and $\theta_j$ is $C_{ij,}$ and its inverse $[\tens{C}^{-1}]_{ij}$. For the MLE of $\bol{\Theta}$, we then minimised
\begin{equation}
    \chi^2(\bol{\Theta})=
    \sum_{i,j=1}^{N_\mathrm{bin}}
    \Big(\gamma_{\mathrm{t},i}(\bol{\Theta})-\gamma_{\mathrm{t},i}\Big)
    [\tens{C}^{-1}]_{ij}
    \Big(\gamma_{\mathrm{t},j}(\bol{\Theta})-\gamma_{\mathrm{t},j}\Big) \;,
\end{equation}
with respect to $\bol{\Theta}$, where $\gamma_{\mathrm{t},i}(\bol{\Theta})$ is the halo model prediction of $\gt(\theta|\bol{\Theta})$ averaged over the size of the $i$th separation bin that corresponds to $\gamma_{\mathrm{t},i}$. We refer to $\bol{\Theta}_\mathrm{mle}$ as the parameter set that minimises $\chi^2(\bol{\Theta})$. Finally, given the MLE $\bol{\Theta}_\mathrm{mle}$, we obtain the MLE of the mean halo mass by the integral
\begin{equation}
    \ev{m}_\mathrm{mle}=
    \int_0^\infty\frac{\mathrm{d} m\,n(m)\,m}{\bar{n}_\mathrm{d}}\;
    \Big( \ev{N_\mathrm{cen}|m}+\ev{N_\mathrm{sat}|m} \Big)\;,
\end{equation}
which has to be evaluated for the HOD of galaxies that is determined by $\bol{\Theta}_\mathrm{mle}$.

\begin{figure}[htbp]
    \centering
    \includegraphics[width=.49\textwidth]{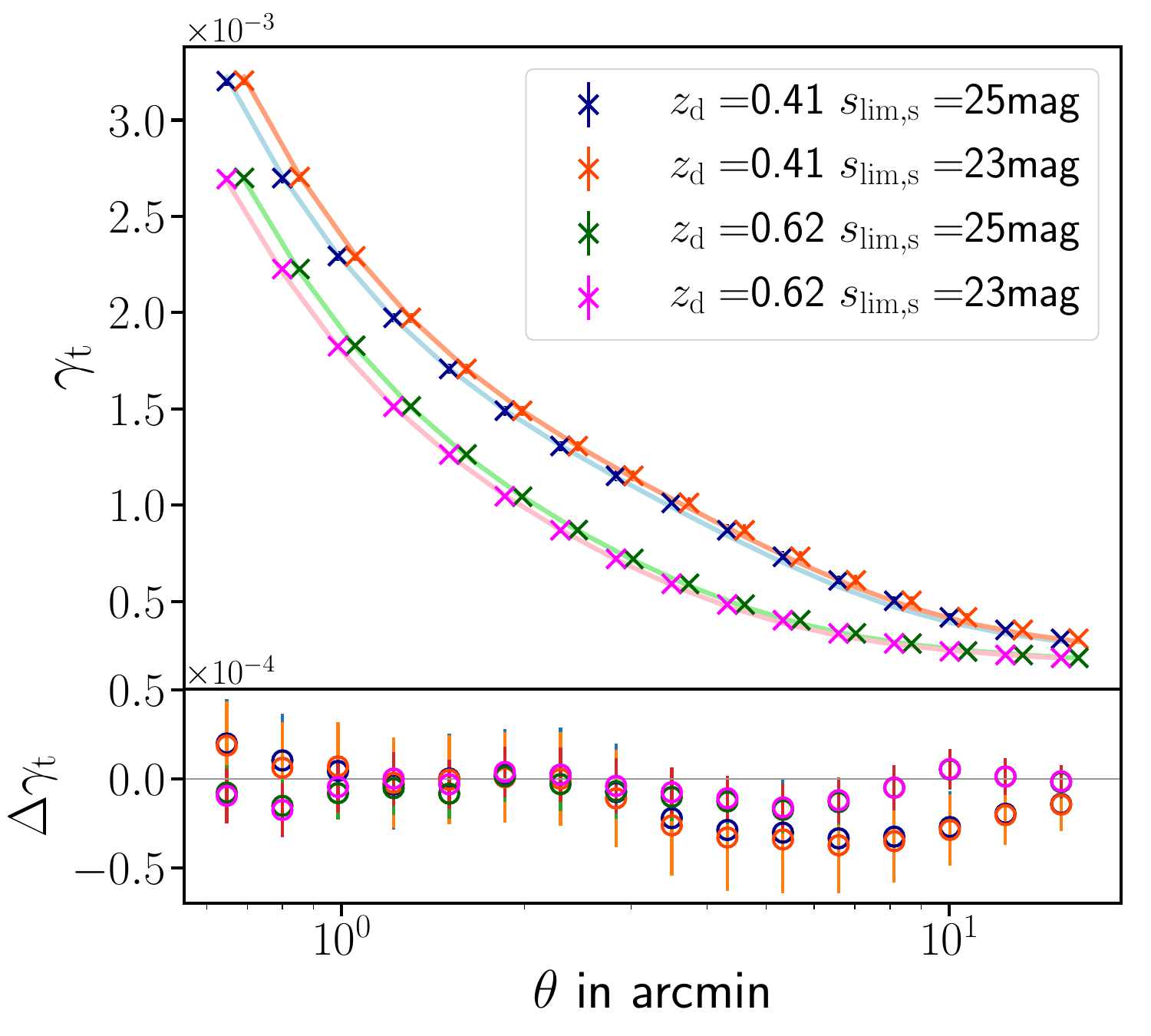}
    \caption{Upper panel: Magnification-corrected shear profiles for two lens redshifts \zd\ with a \(r\)-band magnitude cut of \(22\,\mathrm{mag}\) and two differently selected source galaxies at redshift \(\zs=0.99\) are shown (crosses) as well as their best fit according to the halo model (solid line). Lower panel: Absolute difference \(\Delta \gt\) between data and halo model fit (open circles with the same colour). For each lens redshift, source galaxies were chosen with two limiting magnitudes that have the local slopes \(\as=0.58\) and \(1.75\), respectively. As expected, the shear profile does not depend on the choice of source galaxies. The offset between the red/blue points as well as the magenta/green points is for better visibility only.} 
    \label{pic:shearfit}
\end{figure}
In Fig.\,\ref{pic:shearfit}, we show four examples of tangential shear estimates and their best shear profile fit from the halo model. For the shear profiles, we switched off magnification effects of lens and source galaxies. It can be seen that the fitting procedure works reasonably well. The halo model itself is only an approximation of the inhomogeneous matter distribution in the Universe, and in the presence of our simulated data with almost vanishing Poisson noise, we do not expect the halo model to work perfectly. We further used approximations such as a truncated NFW model and a specific mass-concentration relation, that certainly further limits the accuracy we can obtain with the halo model. The mean relative difference between the data and the model for the 16 bins is \(2.6\%\) for \(\zd=0.41\), and \(1.3\%\) for \(\zd=0.62\). As expected, the shear profiles are almost independent of the flux limit of sources if their redshift is fixed, as can be seen for the red and blue crosses, as well as for the magenta and green crosses. On the other hand, there are three main reasons for the difference between the red/blue and magenta/green shear profiles. The lensing efficiencies \(\Dds/\Ds\) are different for different \zd. Thus, the red/blue shear profile with \(\zd= 0.41\) has a larger lensing efficiency than the magenta/green one with \(\zd=0.62\). Moreover, we observed fixed angular scales, which corresponds to different physical scales. Lastly, the lens galaxy population might evolve between the two redshifts.

Table~\ref{tab:shearfit} accompanies Fig.\,\ref{pic:shearfit} and lists the fitting parameters, the goodness-of-fit values, and the mass estimates for the different shear profiles. The mean halo mass, in contrast to the shear amplitude, is larger for the high-redshift lenses when the same magnitude limit is applied.
\begin{table}[htbp]
    \centering
    \caption{Fitting results for the halo model are shown, with the mean halo mass \(\ev{m}_\mathrm{mle}\), the scatter in host halo mass \(\sigma_{\log m}\), the mass scale where \(50\%\) of halos host a galaxy \(m_\mathrm{th}\), the normalisation factor for the satellite galaxies \(m_1\) and its slope \(\beta\), and the goodness-of-fit value \(\chi^2_\mathrm{red}\) with 12 degrees of freedom.  The lens and source redshifts, and local slopes \ads\ are chosen as in Fig.\,\ref{pic:shearfit}, the source redshift \zs\ is 0.99. The fit values for identical lens redshifts are expected to be very similar.}
    \label{tab:shearfit}
    \begin{tabular}{r|cccc}
         & \multicolumn{4}{c}{plot colour} \\
         & blue & red & green & magenta \\
        \hline\hline
        \zd & 0.41 & 0.41 & 0.62 & 0.62 \\
        \ad & 0.67 & 0.67 & 1.57 & 1.57 \\
        \as & 0.58 & 1.75 & 0.58 & 1.75 \\
        \hline
        \(\ev{m}_\mathrm{mle}\) in \(10^{13}\,\Msun\) & 2.45 & 2.46 & 2.58 & 2.60 \\
        \hline
        \(\sigma_{\log m}\) & 0.28 & 0.28 & 0.28 & 0.30 \\
        \(m_\mathrm{th}\) in \(10^{11}\,\Msun\) & 2.24 & 2.26 & 4.15 & 4.19 \\
        \(\beta\) & 1.08 & 1.09 & 1.06 & 1.05 \\
        \(m_1\) in \(10^{12}\,\Msun\) & 7.70 & 7.68 & 11.54 & 11.92 \\
        \hline
        \(\chi^2_\mathrm{red}\) & 1.52 & 1.84 & 1.10 & 1.25 
    \end{tabular}
\end{table}

We quantify this bias in halo mass for fixed limiting magnitudes of lenses and sources, which means fixed \ads, and fixed redshifts as follows. Using the halo model as described above, we calculated the best mass estimate from the magnification-corrected shear profile. As could be seen in the previous sections, the relative change of the shear profile is typically of the order of a couple of per cent. Therefore, we fixed the scatter in the host halo mass \(\sigma_{\log m}\) and the slope of the mean number of satellite galaxies \(\beta\) to their best fit value in the magnification-corrected case. Then, we only fitted the remaining two parameters \(m_\mathrm{th}\) and \(m_1\) to estimate the mass for the three remaining shear profiles, meaning a shear profile with magnification of the sources only turned on, a profile with magnification of the lenses only turned on, and a shear profile with lens and source magnification turned on. Similar to the fractional shear difference (\ref{eq:sheardiff}), we define the bias of halo-mass estimates, inferred from \gt, by
\eq{
    \label{eq:massdiff}
    \dM = \frac{\ev{m}_\mathrm{mle} - \ev{m}_\mathrm{mle}^\mathrm{nomagn}}{\ev{m}_\mathrm{mle}^\mathrm{nomagn}} \;.
}
\subsection{Numerical results}
\begin{figure}[htbp]
    \centering
    \includegraphics[width=.49\textwidth]{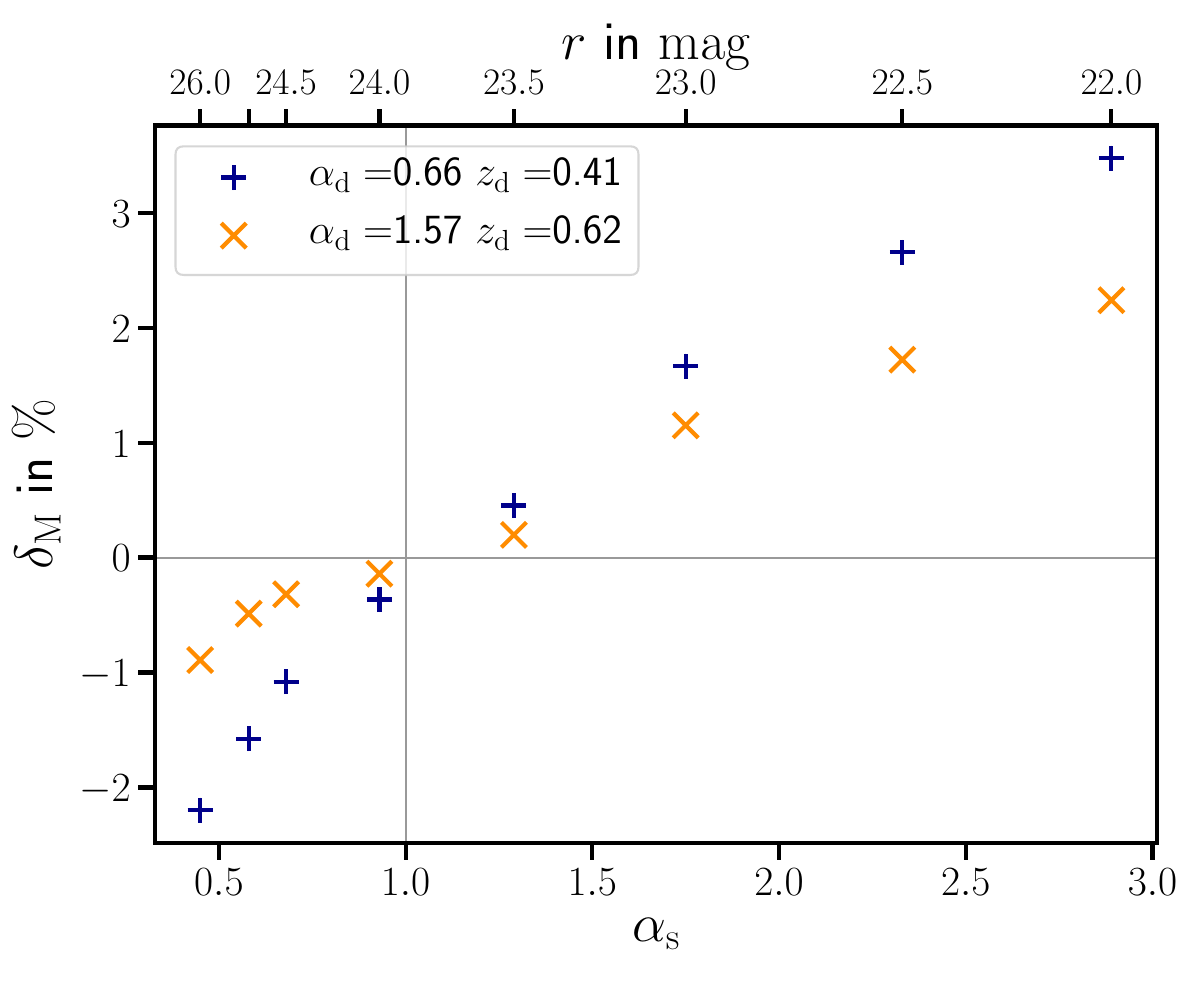}
    \caption{Relative mass bias \dM\ (\ref{eq:massdiff}) for magnified source number counts is shown. Source galaxies are at \(\zs=0.99\) and lenses are chosen according to Table~\ref{tab:magvsa}a with redshifts \(\zd=0.41\) (plus) and 0.62 (cross), for fixed limiting magnitude. The mass bias behaves roughly like the fractional shear difference (cf.~Fig.\,\ref{pic:evolsource}). For local slopes \(\as<1\), the underestimation of mass is stronger than for the shear profile, while for \(\as>1\) the overestimate is similar.}
    \label{pic:evolsourceM}
\end{figure}

\begin{figure}[htbp]
    \centering
    \includegraphics[width=.49\textwidth]{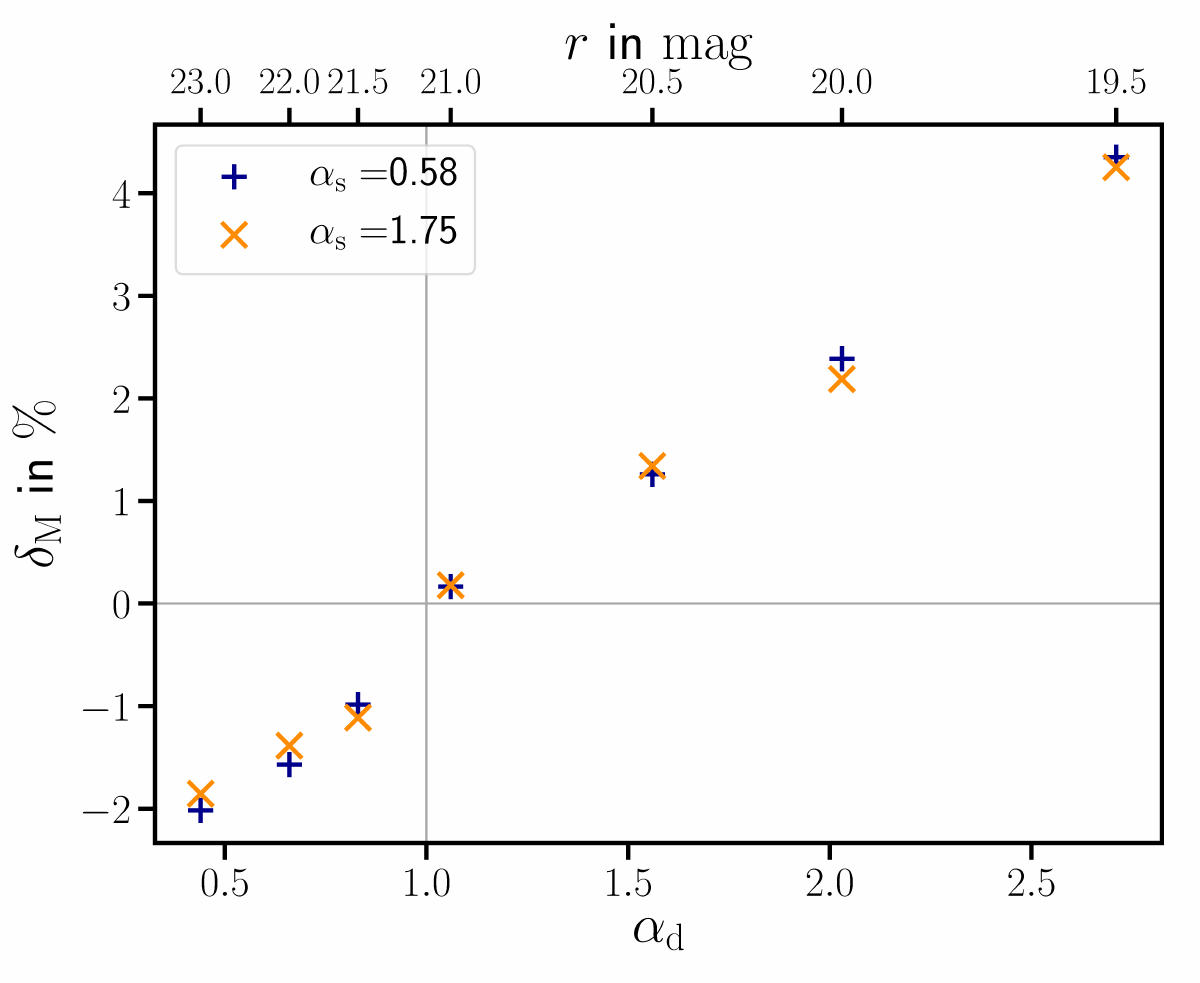}
    \caption{The fractional mass bias \dM\ is shown as a function of local slope \ad\ and \(r\)-band limiting magnitude (cf.~Table~\ref{tab:magvsa}b). The source redshift is \(\zs=0.99\) and lens redshift is \(\zd=0.41\). For \(\ad<1\), \dM\ the mass is biased low, while for \(\ad>1\) the mass is biased high.}
    \label{pic:evollens2M}
\end{figure}

\begin{figure}[htbp]
    \centering
    \includegraphics[width=.49\textwidth]{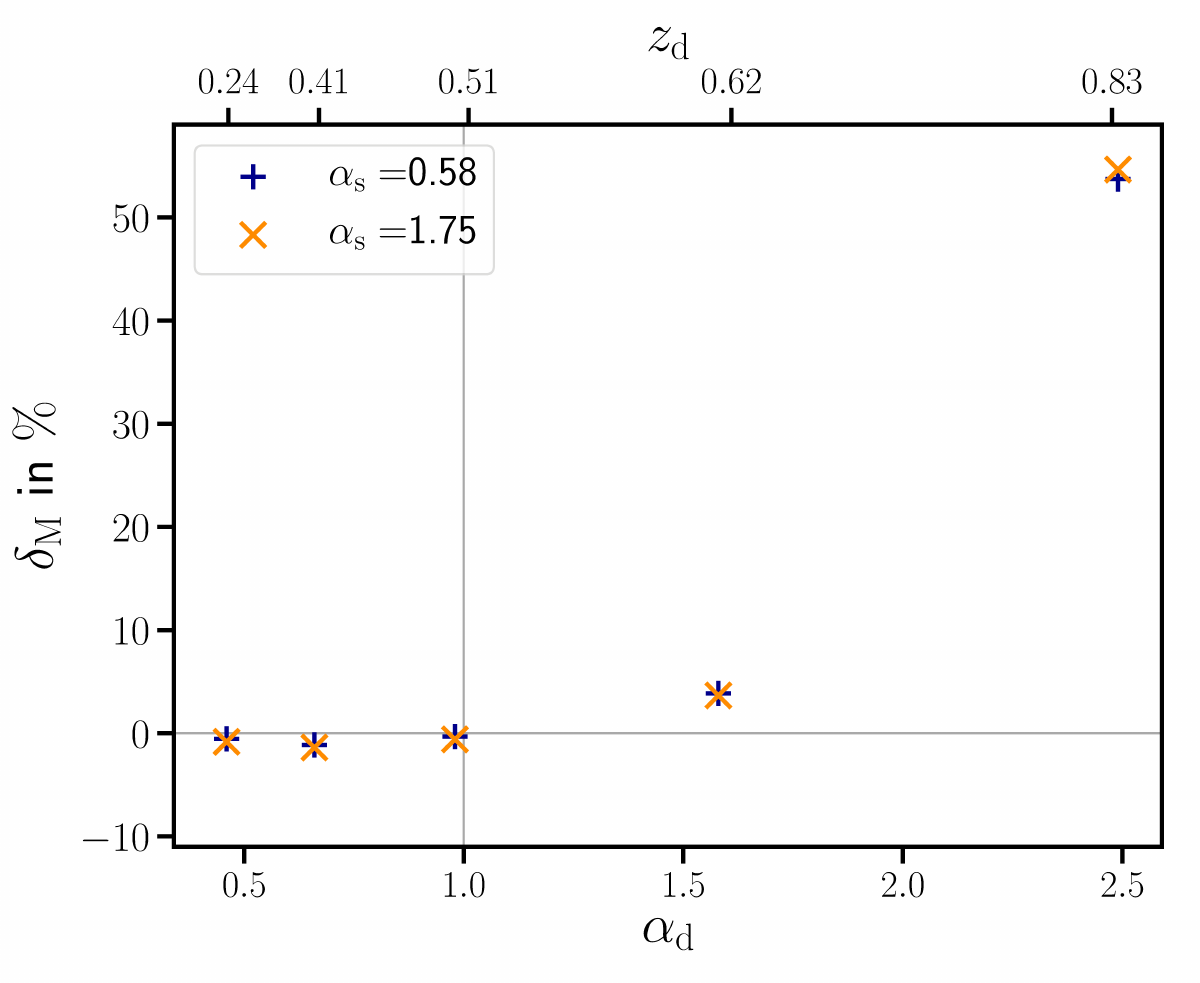}
    \caption{The fractional mass bias \dM\ is shown as a function of local slope \ad\ and lens redshift \zd (cf.~Table~\ref{tab:magvsa}c). We keep the \(r\)-band limiting magnitude for the lenses and the source redshift \(\zs=0.99\) constant. The top axis indicates the respective lens redshifts in a non-linear scaling. Following the trend seen in Fig.\, \ref{pic:evollens}, mass is biased low for \(\ad<1\) and shows large biases for \(\ad>1\).}
    \label{pic:evollensM}
\end{figure}

\begin{figure*}[htbp]
    \centering
    \sidecaption
    \includegraphics[width=.7\textwidth]{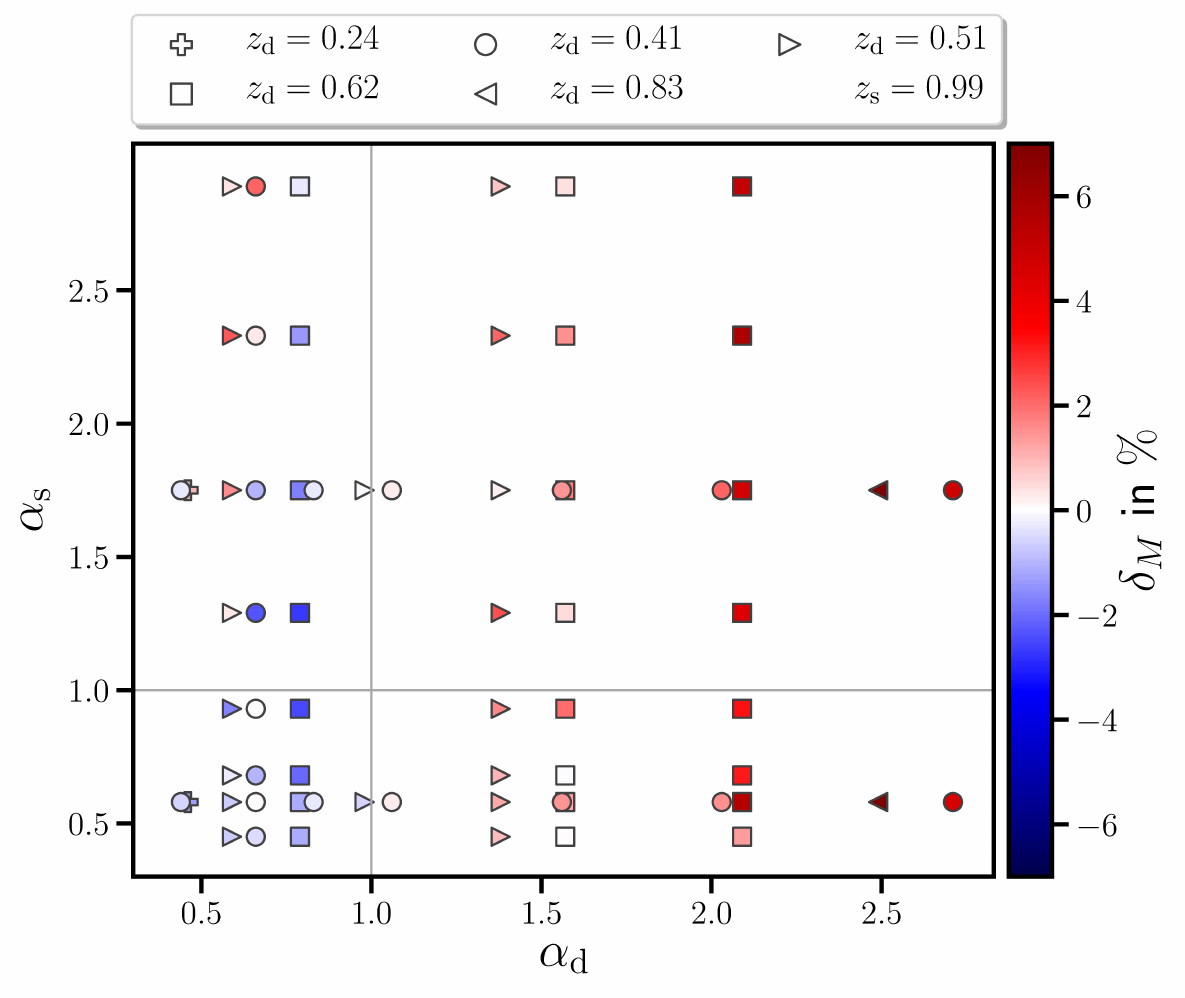}
    \caption{The total magnification bias for halo-mass estimates as a function of limiting magnitudes that yield the local slopes \ads\ and lens redshift \zd\ (type of symbols). The source redshift is the same in all cases, i.e., \(\zs=0.99\). The mass difference \dM\ is shown in colour code, where blue indicates underestimation, red overestimation, and white is an unbiased result. We cut off the colour bar for the largest deviations of \(55\%\) and \(58\%\) for \(\ad=2.41\) and \(\zd=0.83\) for higher contrast in the colour scale. The plot is roughly divided by the vertical line with \(\ad=1\) into mass underestimation for \(\ad<1\) and overestimation \(\ad>1\).}
    \label{pic:massfit}
\end{figure*}

Figures \ref{pic:evolsourceM} to \ref{pic:massfit} show results for the mass bias \dM. Figure \ref{pic:evolsourceM} shows the bias for magnified source galaxy counts and magnification-corrected lens counts.
The lenses have constant limiting magnitude of \(22\,\mathrm{mag}\) in the \(r\)-band and their redshifts are \(\zd=0.41\) and \(0.62\). Source galaxies at redshift \(\zs=0.99\) are selected for several limiting magnitudes (cf.~Table~\ref{tab:magvsa}a). The bias is of the same order of magnitude as the corresponding mean fractional difference of the shear, and \as\ determines whether mass is overestimated or underestimated. The mean halo mass is biased by up to \(3.5\%\).

We then explored the dependencies of the fractional mass bias \dM\ on \ad, while we only considered magnification-corrected sources. 
Firstly, we fixed the lens redshift to \(\zd=0.41\) and altered the lens' limiting magnitude according to Table~\ref{tab:magvsa}b. The result is shown in Fig.\,\ref{pic:evollens2M}. The mass bias is an almost linear function of \ad\ and shows a similar dependence on \ad\ as the fractional shear difference (cf.~Fig.\,\ref{pic:evollens2}). The mass is biased up to \(5\%\) for the largest \ad. Lastly, we calculated \dM\  for various lens redshifts with constant limiting magnitude for the lenses (see Table~\ref{tab:magvsa}c) and the same two \as\ as before, which is shown in Fig.\,\ref{pic:evollensM}. 
The mass bias shows a strong redshift dependence, where the bias increases from a couple of per cent to a mass overestimate of \(55\%\). 

To explore the observationally relevant case, we compared halo-mass estimates with and without magnification for both lenses and sources. Figure~\ref{pic:massfit} contains all different \ads-\zd\ combinations with constant \(\zs=0.99\) from the Tables~\ref{tab:magvsa}a to \ref{tab:magvsa}c, plus some additional combinations. 
It shows \dM\ as defined in Eq.\,(\ref{eq:massdiff}) as a function of the local slopes \ads\ on the \(x\)- and \(y\)-axis. Red values indicate an overestimation of the mass, where we cut off the colour bar for the highest values for better visibility. The blue values show an underestimation, while white values mean no bias in the mass estimate. The upper-right quadrant shows consistently red values since \(\ads>1\), while the lower-left quadrant has \(\ads<1\) and is consistently blue, as expected from theoretical considerations. In most of the other cases, \ad\ seems to be the decisive factor for the sign of the mass bias, for \(\ad<1\) leading to an underestimation and for \(\ad>1\) leading to an overestimation. The slope \as\ influences the total value of the mass bias. Besides, even if two points are close in the \ad-\as\ parameter space, they do not exhibit the same colour, meaning mass bias, due to their different \zd.
\section{\label{sec:Sc7}Discussion \& conclusions}
%
%
In this paper, we quantified the impact of magnification on the GGL signals and halo-mass estimates. Magnification changes the observed galaxy number counts on the sky, which has an impact on the measured tangential shear profiles. It is important to note that magnification affects lens galaxies as well as source galaxies. Analyses of tangential shear profiles, such as estimates of the excess mass density profiles or halo mass, are therefore biased if they ignore lens or source magnification.
%
%
Our estimates with ray-tracing simulations and synthetic galaxy populations show 
that the number count slopes of sources and lenses are the most important quantities that determine the relative strength of the bias. While the analytical estimate for the bias on the GGL profile caused by magnification of lenses was known before, to our knowledge the one caused by magnification of the source population has not previously been derived.

How magnification affects tangential shear profiles can be seen in Figs.~\ref{pic:shearprofs} to \ref{pic:evollens}, where we varied the local slope of lens and sources galaxies \ads\ and the lens redshift \zd. We studied the magnification effect from lenses and sources individually and compared them to our analytic estimates. The latter are leading-order estimates and describe the numerical results very well in most cases. One of the surprising results, shown in Fig.\,\ref{pic:sheardiff_source}, is that the weak lensing approximation for the impact of magnification on source counts can significantly fail if the count slope is steep. Hence, the validity of the commonly used approximation for the number density of sources \[\ns(\btheta, >s) \approx n_\mathrm{s0}(>s)\left[ 1 + 2(\as-1) \kappa(\btheta) \right]\] needs to be checked, depending on its application.


Besides confirming that the shear signal is reduced if the slopes \ad\ and \as\ are \(<1\) and enhanced for \(\ads>1\), we find: For fixed redshifts \zds,\, the change of the tangential shear estimate depends solely on \ads. For fixed source redshifts \zs, the impact of magnification of sources decreases for larger \zd\ and smaller lens-source separation \Dds\ (cf.~Fig.\,\ref{pic:evolsource}). Furthermore, the relative importance of magnification of lenses increases with \zd\ and rises sharply as \zd\ approaches \zs. The relative change from a biased to an unbiased shear profile is a function of angular separation $\theta$, and the mean change is typically a few per cent. However, if the redshift differences between sources and lenses become small, the effect can be considerably larger (cf.~Fig.\,\ref{pic:evollens}).

For practical applications, the change in the shear signal by magnification is described by the sum of the individual effects of source magnification and the one from lens magnification. The impact of magnification depends on the galaxy's luminosity function, the limiting magnitudes, and the redshifts \zds, but also the separation $\theta$.

The relative mass bias \dM\ on the average mass of a lens parent-halo inherits all trends seen in the shear profiles with magnified lenses and magnified sources (see Figs.~\ref{pic:evolsourceM} to \ref{pic:massfit}), meaning \(\dM>0\) for \(\ads>0\) and vice versa. Medium-redshift galaxies show a mass bias of less than \(10\%\), and the higher the lens redshift, the stronger the mass bias with up to \(58\%\) for \(\zd=0.83\) and \(\zs=0.99\). The bias \dM\ is of the same order of magnitude as the relative change of the shear profile \tgt. The particular relation, however, between \tgt\ and the mass bias \dM\ is quite complicated. First and foremost, the amplitude of the measured shear signal determines the underlying halo mass. However, magnification changes the scale dependence of the shear signal. In the halo model, this translates to a different behaviour of the one- and two-halo term, which affects the mass estimate in a highly non-linear way. Another minor effect is that the true mean halo mass of lens galaxies affected by magnification probably differs from the true mean halo mass of unmagnified lenses due to an expected correlation between mass and luminosity. For example, for the highest \zd\ considered in this work, the mean masses differ by \(0.16\%\). Furthermore, we fixed the observed angular scales on the sky, which relate to different physical scales at different redshifts; thus, the relative contribution of the two-halo term on the shear signal grows with redshift, which makes a comparison of mass biases from different lens redshifts more complicated. In general, when we allow for magnification effects both in source and lens galaxies, meaning, the case for real observations, the sign of the mass bias is in most cases determined by the value of \ad. The only exceptions shown in Fig.\,\ref{pic:massfit} are two cases, where either the lens redshift is low or the local slope \as\ is very high; in such cases, the sign of the mass bias is not easily predicted and must be studied case by case.

We also considered a GGL estimate using the flux-limits and redshift distributions as given from the combined data of KiDS+VIKING \citep{hildebrandt2020} and GAMA \citep{Driver2011}. KiDS and VIKING are partner surveys that probe the optical and near-infrared sky to obtain high-resolution, wide-field shape and redshift information from galaxies. GAMA is a spectroscopic, flux-limited survey with a partly overlapping footprint in the KiDS+VIKING area. In Appendix \ref{sec:forecast}, we present a detailed description of the input parameters we use for this estimate. We list the results in Table \ref{tab:kids}. A lens galaxy sample at redshift \(\zd=0.21\) shows relative changes in the shear profile due to magnification effects of less than one per cent, and relative mass bias of \(\approx 3\%\), while GAMA's highest lens sample at \(\zd=0.36\) shows that a magnification correction changes the shear profile by \(\approx 2\%\) and the relative halo mass bias by \(8\%\). Although our estimate used simplifying assumptions, especially for the source population, for example, no catastrophic outliers in the redshift distribution and no further selection criteria than a cut in magnitude, we conclude that magnification effects must be carefully considered in current and future surveys.

In this paper, we assume that lenses and sources form a flux-limited sample. While this assumption may be a realistic one for lens galaxies (e.g. galaxy redshift surveys frequently start from a flux-limited photometric sample), it is less the case for source galaxies. Sources in weak lensing studies have rather complicated selection criteria, not merely based on flux, but also on size and signal-to-noise ratio, for example. Therefore, our quantitative analysis may not apply directly to observational surveys. Besides, source galaxies typically enter a weak lensing catalogue with a weight that characterises the accuracy of the corresponding shear estimate. We ignored any such weighting scheme in our processing, but it may be relevant, since the weight of an object is also expected to depend on magnitude and size, and is thus affected by magnification.

While the relative amplitude of the bias caused by magnification is modest in most cases, and probably smaller than the uncertainties from shape noise and sample variance in previous surveys, future surveys like \textit{Euclid} or LSST have such improved statistical power that magnification effects must be accounted for in the quantitative analysis of GGL.
\begin{acknowledgements}
    The authors would like to thank the anonymous referee for his constructive and helpful comments that improved the quality of this paper. 
    Part of this work was supported by the \emph{Deut\-sche For\-schungs\-ge\-mein\-schaft, DFG\/} project number SCHN\,342/13-1.
    Sandra Unruh is a member of the International Max Planck Research School (IMPRS) for Astronomy and Astrophysics at the Universities of Bonn and Cologne.
    Stefan Hilbert acknowledges support by the DFG cluster of excellence \lq{}Origin and Structure of the Universe\rq{} (\href{http://www.universe-cluster.de}{\texttt{www.universe-cluster.de}}).
\end{acknowledgements}

\bibliography{Magnbias_massbib}
\bibliographystyle{aa}

\appendix

\section{Mock data}
\label{sec:mockdata}

\subsection{Millennium Simulation data}

To study magnification effects in GGL, we make use of ray-tracing results  through the Millennium Simulation \citep[MS,][]{Springel2005}, which is an \(N\)-body simulation of \(2160^3\) dark matter particles. Each particle has a mass of \(8.6 \times 10^8 \, \Msun\) that is confined to a cube with side length of \(500\,h^{-1} \mathrm{Mpc}\) and with periodic boundary conditions. The underlying cosmology is a flat \(\Lambda\)CDM model with a matter density parameter of \(\Omega_\mathrm{m} =0.25\), a baryon density parameter of \(\Omega_\mathrm{b} = 0.045\), a dimensionless Hubble parameter of \(h =0.73\), a tilt of the primordial power spectrum of \(n=1\), and a variance of matter fluctuations on a scale of \(8\,h^{-1}\,\mathrm{Mpc}\) extrapolated from a linear power spectrum of \(\sigma_8 = 0.9\). This cosmology is based on combined results of 2dFGRS \citep{Colless2001} and first-year WMAP data \citep{Spergel2003}.

The ray-tracing results are based on a multiple-lens-plane algorithm in 64 light cones constructed from 37 snapshots between redshifts \(z=0\) to \(z=3.06\), each covering a \(4\times 4 \, \degt^2\) field of view. For more information about the ray tracing, the reader is kindly referred to \citet{Hilbert2009}. An important aspect of the algorithm is that the galaxy-matter correlation is preserved. The ray-tracing results contain the Jacobians \(\mathcal{A}\) on a \(N_\mathrm{pix} = 4096^2\) pixel grid, which corresponds to a resolution of \(3.5\,\arcsect\) per pixel. From this, we calculated shear and magnification on a pixel grid.

We also used a catalogue of galaxies based on a semi-analytic galaxy-formation model by \citet{Henriques2015}. This catalogue matches the GGL and galaxy-galaxy-galaxy lensing signal from CFHTLenS \citep{Saghiha2017}. The galaxies are listed for each redshift snapshot with various properties, for example, (magnified) flux in various filters and positions, which allows for selection according to chosen magnitude limits. However, the galaxy positions are not confined to a grid as is the Jacobi information. Thus, we shifted all selected galaxies to their nearest grid point. Therefore, analyses that are close to the centre of the galaxy suffer from discretisation effects on scales comparable to the pixel size.

\subsection{Obtaining a tangential shear estimate}

To extract the shear signal averaged over many lenses, a fast Fourier transform (FFT) is employed. In order to do so, we first defined lens and source number density on a grid by
\eq{
    \label{eq:numdensity}
    n_\mathrm{d,s} (\btheta) =
    \sum^{N_\mathrm{d,s}}_{i=1} \deltaK (\btheta - \btheta^{(i)}_\mathrm{d,s})\;,
}
with \(\deltaK\) being one if \(\btheta = \btheta^{(i)}_\mathrm{d,s,}\) and zero otherwise. The number of lenses and sources is \(N_\mathrm{d,s}\), and \(\btheta^{(i)}_\mathrm{d,s}\) are the positions of lenses and sources, respectively. Furthermore, we define the shear field of the sources on the grid by
\begin{equation}
    \label{eq:source_shear}
    \gamma_\mathrm{s} (\btheta) =
        \sum^{N_\mathrm{s}}_{i=1} \gamma(\btheta, \zs) \, \deltaK (\btheta - \btheta^{(i)}_\mathrm{s}) \;.
\end{equation}

Then, the tangential shear estimator~\eqref{eq:ggl_gtx_estimator} can be expressed as
\begin{equation}
    \label{eq:grid_gt_estimator}
    \est{\gamma}_{\mathrm{t}} (\theta) = 
    -\Re
    \frac{
    \sum_{\btheta'} \Delta(\theta, |\btheta'|)\, \btheta'^*/\btheta'\;
    \sum_{\btheta''} \nd (\btheta'') \, \gs(\btheta'' + \btheta')
    }{
    \sum_{\btheta'} \Delta(\theta, |\btheta'|)\;
    \sum_{\btheta''} \nd (\btheta'') \, \ns(\btheta'' + \btheta')
    } \;,
\end{equation}
where the sums over $\btheta'$ and $\btheta''$ extend over the whole grid. The equivalence of Eqs.\,(\ref{eq:grid_gt_estimator}) and (\ref{eq:ggl_gtx_estimator}) can be verified by inserting the definitions of $n_\mathrm{d,s}$ and $\gamma_\mathrm{s}$ into the former. The sum over $\btheta''$ in the denominator calculates for each \(\theta'\) the number of lens-source pairs, ensuring that (\ref{eq:grid_gt_estimator}) is not affected by a potential masking or inhomogeneous survey areas. Further, the sums over $\btheta''$ in the numerator and denominator of the GGL estimator~\eqref{eq:grid_gt_estimator} are convolutions. Thus, the convolution theorem can be applied to these sums. If \(\mathcal{F}\{f\}\) is the Fourier transform of a function \(f\) and \(\mathcal{F}^{-1}\) the inverse Fourier transform, then we can rewrite the estimator~(\ref{eq:grid_gt_estimator}) as
\begin{equation}
    \label{eq:grid_gt_estimator_ft}
    \est{\gamma}_{\mathrm{t}} (\theta) = 
    -\Re 
    \frac{
    \sum_{\btheta'} \Delta(\theta, |\btheta'|)\, \btheta'^*/\btheta'\;
    \ift{\ft{\nd} \ft{\gs}}(\btheta')
    }{
    \sum_{\btheta'} \Delta(\theta, |\btheta'|)\;
    \ift{\ft{\nd} \ft{\ns}}(\btheta')
    } \;,
\end{equation}
which can be readily solved by an FFT method. For this, we employed routines from the FFTW library by \citet{FFTW} in our code.

An FFT implicitly assumes periodic boundary conditions, which introduces a bias to the averaged shear data and, thus, must be mitigated for. We can restrict the selection of lenses to the inner \((4^\circ-2\theta_\mathrm{out})^2\) of the field, where \(\theta_\mathrm{out}\) is the maximum separation from the lens that we considered. However, for a \(\theta_\mathrm{out} \approx 20',\) we already lose approximately \(30\%\) of the lens galaxies. Alternatively, we  employed a zero-padding method in which we increased the FFT-area to \((4^\circ+\theta_\mathrm{out})^2\) and filled the added space with zeros. In this case, we used all available lenses with the cost of slightly increased computational time and the gain of a less-noisy shear profile. For the whole \(64\times 16\,\degt^2\), our FFT-based code needs a CPU time of \(823 \, \mathrm{s}\), independent of the number of sources and lenses. We compared the performance of our method to the publicly available \texttt{athena} tree-code \citep{Kilbinger2014}. In contrast to the FFT method, the computation time of \texttt{athena} is enhanced with the number of lens-source pairs. We adjusted the settings to our survey parameters while leaving the parameter that sets the accuracy of the tree-code, meaning the open-angle threshold, to its pre-set value. For \(10^7\) sources and lenses at \(\zd=0.41\) with limiting magnitudes \(19.5\), \(22\), and \(29\,\mathrm{mag}\), \texttt{athena} performs with a CPU time of \(889 \, \mathrm{s}\), \(1167 \, \mathrm{s}\), and \(2043 \, \mathrm{s}\), respectively.

\subsection{Estimating the impact of the magnification bias on a KiDS+VIKING+GAMA-like survey}
\label{sec:forecast}

We considered GAMA-like lens galaxies with narrow redshift distributions at \(\zd=0.21\) and \(\zd=0.36\) with a flux limit of \(19.8\,\mathrm{mag}\). Using simulated data from the semi-analytic galaxy-formation model by \citet{Henriques2015}, we obtain local slopes of \(\ad=0.85\) and \(\ad=2.11\), respectively. To keep the statistical error low, we still consider the data from the whole simulated area of $64\times 16\,\degt^2$. To mimic the KiDS+VIKING-like source population, we matched the last three bins of the best-estimated redshift distribution to Millennium data from \(0.51\le z_\mathrm{s,Mil}<1.28\). We then calculated a weighted shear map from the simulated data, the final source galaxy distribution has a local slope of \(\as=0.51\) for a limiting magnitude of \(25\,\mathrm{mag}\). We repeated our analysis from Sects.\,\ref{sec:magnification_effects_on_background_galaxies}, \ref{sec:magnification_effects_on_foreground_galaxies}, and \ref{sec:magnification_bias_in_halo_mass_estimates}, and obtain the results listed in Table \ref{tab:kids}. The results are quantitatively comparable to the results from Figs.\,\ref{pic:evolsource} and \ref{pic:evollens2}.
\begin{table}[htbp]
    \centering
    \caption{Impact of the magnification bias on a KiDS+VIKING+GAMA-like survey. The superscript `d' indicates the magnification bias caused only by the lens galaxies, while `s' stands for the source galaxies.}
    \label{tab:kids}
    \begin{tabular}{rrr}
        \zd & \(0.21\) & \(0.36\) \\
        \ad & \(0.85\) & \(2.11\) \\
        \as & \(0.51\) & \(0.51\) \\
        \({\tgt}^\mathrm{d}\) & \(-0.09\%\) & \(1.67\%\) \\
        \({\tgt}^\mathrm{s}\) & \(-0.88\%\) & \(-0.94\%\) \\
        \dM & \(-2.84\%\) & \(-8.26\%\) 
    \end{tabular}
\end{table}

A KiDS+VIKING+GAMA-like survey is moderately affected. We stress that this estimate has been made with simplified assumptions, for example, there are no catastrophic outliers in the redshift distribution and no selection criteria for source galaxies other than a magnitude cut.


\end{document}